\documentclass[preprint,aps,floatfix,nofootinbib,a4paper, superscriptaddress]{revtex4-1}

\usepackage{amsmath, amssymb, amsfonts, amsthm, latexsym, epsfig, mathrsfs, xcolor, bbm, slashed, float}

\usepackage[all]{xy}

\usepackage[inline]{enumitem}

\usepackage{setspace}
\usepackage[marginal, multiple]{footmisc}

\usepackage[utf8]{inputenc}
\usepackage[T1]{fontenc}
\usepackage{lmodern}

\usepackage{microtype}

\usepackage[colorlinks, allcolors=blue!70!black, linktocpage]{hyperref}

\numberwithin{equation}{section}

\usepackage{cleveref}

\let\OLDthebibliography\thebibliography
\renewcommand\thebibliography[1]{%
	\setstretch{1.079} 
	\OLDthebibliography{#1}%
	\small %
	\setlength{\itemsep}{0.2\baselineskip} 
}

\let\OLDfootnote\footnote
\renewcommand\footnote[1]{%
	\setlength{\footnotesep}{0.75\baselineskip}%
	{\footnotesize \OLDfootnote{#1}}%
}

\setlist[enumerate]{noitemsep, label=(\arabic*), ref=(\arabic*)}

\renewcommand\thesection{\arabic{section}}
\renewcommand\thesubsection{\arabic{subsection}}

\makeatletter
\def\p@subsection{\thesection.}
\def\p@subsubsection{\thesection.\thesubsection.}
\makeatother 

\theoremstyle{plain}

\newtheorem{lemma}{Lemma}[section]
\newtheorem{prop}{Proposition}[section]

\theoremstyle{definition}

\theoremstyle{remark}
\newtheorem{remark}{Remark}[section]

\crefname{equation}{Eq.}{Eqs.}
\creflabelformat{equation}{#2#1#3}

\crefname{section}{Sec.}{Secs.}
\crefname{appendix}{Appendix}{Appendices}
\crefname{figure}{Fig.}{Figs.}

\crefname{definition}{Def.}{Defs.}
\crefname{prop}{Prop.}{Props.}
\crefname{lemma}{Lemma}{Lemmas}
\crefname{corollary}{Cor.}{Cors.}
\crefname{thm}{Theorem}{Theorems}
\crefname{remark}{Remark}{Remarks}

\crefname{ass}{Assumptions}{Assumptions}
\crefname{property}{Properties}{Properties}

\newcommand{\be}{\begin{equation}}
\newcommand{\ee}{\end{equation}}

\newcommand{\lb}{\left}
\newcommand{\rb}{\right}

\newcommand{\mc}{\mathcal}

\newcommand{\ms}{\mathscr}

\newcommand{\bb}{\mathbb}

\newcommand{\eqsp}{\, ,\quad} 

\newcommand{\inter}{\cap} 

\newcommand{\abs}[1]{\lb\vert\, #1 \,\rb\vert}		

\newcommand{\Lie}{\pounds} 
\newcommand{\defn}{\mathrel{\mathop:}=} 

\newcommand{\adj}[1]{{#1}^\dagger} 

\renewcommand{\Re}{{\rm Re\,}}
\renewcommand{\Im}{{\rm Im\,}}

\DeclareMathOperator{\thorn}{\text{\rm \th}}
\let\eth\relax
\DeclareMathOperator{\eth}{\text{\rm \dh}}

\newcommand{\wt}{\circeq}

\renewcommand{\bar}{\overline}

\newcommand{\nfrac}[2]{{{}^#1\!\!/\!_#2}}

\newcommand{\rsub}[1]{\scriptscriptstyle{\rm #1}}

\begin{document}

\setstretch{1.2}


\title{Canonical Energy and Hertz Potentials for Perturbations of Schwarzschild Spacetime}

\author{Kartik Prabhu}\email{kartikprabhu@cornell.edu}
\affiliation{Cornell Laboratory for Accelerator-based Sciences and Education (CLASSE),\\ Cornell University, Ithaca, NY 14853, USA}
\author{Robert M. Wald}\email{rmwa@uchicago.edu}
\affiliation{Enrico Fermi Institute and Department of Physics, The University of Chicago, Chicago, IL 60637, USA}

\begin{abstract}

Canonical energy is a valuable tool for analyzing the linear stability of black hole spacetimes; positivity of canonical energy for all perturbations implies mode stability, whereas the failure of positivity for any perturbation implies instability. Nevertheless, even in the case of \(4\)-dimensional Schwarzschild spacetime --- which is known to be stable --- manifest positivity of the canonical energy is difficult to establish, due to the presence of constraints on the initial data as well as the gauge dependence of the canonical energy integrand. Consideration of perturbations generated by a Hertz potential would appear to be a promising way to improve this situation, since the constraints and gauge dependence are eliminated when the canonical energy is expressed in terms of the Hertz potential. We prove that the canonical energy of a metric perturbation of Schwarzschild that is generated by a Hertz potential is positive. We relate the energy quantity arising in the linear stability proof of Dafermos, Holzegel and Rodnianski (DHR) to the canonical energy of an associated metric perturbation generated by a Hertz potential. We also relate the Regge-Wheeler variable of DHR to the ordinary Regge-Wheeler and twist potential variables of the associated perturbation. Since the Hertz potential formalism can be generalized to a Kerr black hole, our results may be useful for the analysis of the linear stability of Kerr.
\end{abstract}

\maketitle
\tableofcontents

\section{Introduction}\label{sec:intro}

Determining the stability of black hole solutions to Einstein equation is a long-standing problem in general relativity. In the past few years, considerable progress has been made in the analysis of the stability of Schwarzschild and Kerr black holes. A complete proof of linear stability, including decay, of gravitational perturbations of a \(4\)-dimensional Schwarzschild black hole has been given by Dafermos, Holzegel and Rodnianski (DHR) \cite{DHR}. The method of \cite{DHR} relied on finding a tensorial variable constructed out of the perturbed Weyl tensor satisfying a Regge-Wheeler-type equation for which boundedness and decay could be shown by adapting methods \cite{Daf-Rod-lec} used for scalar fields. The analysis of \cite{DHR} was recently generalised to obtain decay results for perturbed Weyl tensor components for Kerr spacetimes \cite{Ma, DHR-kerr-slow, DHR-kerr}. Recently, the non-linear stability of Schwarzschild for perturbations with a hypersurface orthogonal axial Killing field was shown by Klainerman and Szeftel \cite{KS-Sch-pol}.

Nevertheless, the general stability problem remains open, and it is of interest to develop methods that may yield insights into the analysis of black hole stability.
A general approach to analyze linear stability to axisymmetric\footnote{The restriction to axisymmetric perturbations applies only to rotating black holes. For a static black hole such as Schwarzschild, the axisymmetric restriction may be dropped.} perturbations is given by the canonical energy method of Hollands and Wald \cite{HW-stab}. Positivity of canonical energy immediately implies mode stability \cite{HW-stab}, whereas failure of positivity implies that there exist perturbations that grow exponentially in time \cite{PW}. However, the expression for the canonical energy of gravitational perturbations is quite complicated, and it is not straightforward to analyze its positivity even in spacetimes as simple as Schwarzschild. 

To see the difficulty in analyzing positivity more explicitly, let $\Sigma$ be a Cauchy surface for the black hole exterior that is invariant under the $t$-$\phi$ reflection isometry \cite{SW-tphi}. Let $q_{ab}$ denote the perturbed induced metric on $\Sigma$ and let $p_{ab}$ denote the perturbed canonical momentum. 
The general expression for the canonical energy \(\ms E\) in terms of \((p_{ab}, q_{ab})\) can be found in Eq.~86 of \cite{HW-stab}; this expression occupies half of a journal page. However, this expression simplifies considerably in the case of a static black hole. We obtain,
\be
\ms E = \ms K + \ms U \, .
\ee
Here the \emph{kinetic energy} \(\ms K\) is given by
\be\label{eq:KE-defn}
	\ms K = \tfrac{1}{16\pi} \int_\Sigma 2 N~ \lb[ p_{ab} p^{ab} - \tfrac{1}{2} p^2 \rb]
\ee
where the natural background volume element is understood in the integration over \(\Sigma\). The \emph{potential energy} \(\ms U\) is given by\footnote{The manipulations needed to bring the potential energy expression into this form can be found in Sec.~6.1 of \cite{PSW}, where similar computations were performed with fluid matter.}
\be\label{eq:PE-defn}
	\ms U = -\tfrac{1}{16\pi} \int_\Sigma N \lb[ \tfrac{1}{2} D_c q_{ab} D^cq^{ab} - D_c q_{ab} D^a q^{cb} - \tfrac{3}{2} D_a q D^a q + 2 D^b q D^aq_{ab} \rb] 
\ee
Note that \cref{eq:PE-defn} has an overall negative sign (as compared to the formulas found in \cite{HW-stab,PSW,PW}) because we work with a negative-definite spatial metric, since we will be using the Newman-Penrose (NP) and Geroch-Held-Penrose (GHP) formalisms, for which the $(+,-,-,-)$ signature is standard. (This sign change of the spatial metric affects the sign of all terms with an odd total number of metric contractions.) The perturbed initial data \((p_{ab}, q_{ab})\) appearing in \cref{eq:KE-defn,eq:PE-defn} are not free but must satisfy the linearized constraint equations
\begin{subequations}\label{eq:lin-constraints}\begin{align}
	D^ap_{ab} & = 0 \label{eq:lin-mom-constraint}\\
	D^a D_a q - D^a D^b q_{ab} + R^{\rsub{(3)}}_{ab} q^{ab} & = 0 \label{eq:lin-Ham-constraint}
\end{align}\end{subequations}
where \(R^{\rsub{(3)}}_{ab}\) the Ricci tensor on \(\Sigma\).

We have previously shown that the kinetic energy is positive-definite, i.e., \(\ms K \geq 0 \) with equality only when $p_{ab}$ is pure gauge (Theorem~1 \cite{PW}); indeed, this positivity result for kinetic energy can be generalized to the general, stationary-axisymmetric case. However, we have not been able to directly establish the positivity of the potential energy \(\ms U\),  \cref{eq:PE-defn}, even for the Schwarzschild background.\footnote{As will be discussed further in \cref{sec:positive-can-energy}, using the hypersurface orthogonality of the axial Killing field of Schwarzschild, metric perturbations of Schwarzschild can be further decomposed into axial and polar parts. The axial part automatically satisfies the Hamiltonian constraint \cref{eq:lin-Ham-constraint}, and it is not difficult to show that its contribution to the potential energy is positive. However, the polar part is subject to the constraint \cref{eq:lin-Ham-constraint}, and we have not been able to directly establish its positivity.} The two main obstacles to showing positivity of \(\ms U\) appear to be the following: 
\begin{enumerate*}
	\item The variable \(q_{ab}\) is not free data but is subject to the linearized Hamiltonian constraint \cref{eq:lin-Ham-constraint}.
	\item Even though \(\ms U\) is gauge-invariant, the integrand in \cref{eq:PE-defn} is not.
\end{enumerate*}
Thus, it would seem that to show positivity of \(\ms U\) one would need to utilize the constraint equations effectively and make a suitable choice of gauge (as was done for the proof of positivity of \(\ms K\) in \cite{PW}). However, we have not, as yet, found a way to do this.

A promising strategy to prove positivity would be rewrite the canonical energy in terms of unconstrained variables. In \(4\)-dimensions, for the case of algebraically special spacetimes such as Schwarzschild and Kerr, a possible choice of such variables are the \emph{Hertz-Bromowich-Debye-Whittaker-Penrose potentials} (henceforth Hertz potentials)  \cite{CK-grav,Chrz,Stewart,Wald-potential}. The Hertz potentials, which solve the Teukolsky equation \cite{Teu-more}, can be used to generate (complex) metric perturbations that solve the Einstein equation. The initial data corresponding to the Hertz potentials is unconstrained, and one could attempt to prove positivity of canonical energy for perturbations generated by a Hertz potential.\footnote{The suggestion of expressing canonical energy in terms of Hertz potentials was first made to us by Lars Andersson. A generalization of the Hertz potential formalism combined with the canonical energy method was used by \cite{HI-ex-stab} to give a criteria for instablilites of extremal black holes in higher dimensions.}

The main purpose of this paper is to show that for perturbations of Schwarzschild generated by a Hertz potential, the canonical energy is indeed positive. Since the Hertz potential formalism can be straightforwardly extended to Kerr spacetime, it is possible that the methods of this paper may be useful in the analysis of the linear stability of Kerr.

We begin by reviewing properties of the Schwarzschild spacetime in the Carter null frame in \cref{sec:Sch-bg}. In \cref{sec:hertz} we define the Hertz potentials and the metric perturbations generated by them, following \cite{Wald-potential}. We show (but do not provide a complete proof) that --- apart from the trivial perturbations to other Kerr black holes --- any perturbation of Schwarzschild of compact support can be arbitrarily well approximated in an $L^2$ norm by a perturbation obtained from a Hertz potential plus a gauge transformation. In \cref{sec:positive-can-energy} we show that on Schwarzchild spacetime, the canonical energy of perturbations generated by a Hertz potential is positive. In \cref{sec:DHR-compare} we relate the energy quantity underlying the analysis of DHR \cite{DHR} to the canonical energy of an associated perturbation generated by a Hertz potential. We relate the variable used by DHR \cite{DHR} to the Regge-Wheeler and twist potential variables of the associated perturbation in \cref{sec:RW-twist-compare}. In \cref{sec:EM} we carry out the analogous analysis for electromagnetic perturbations of Schwarzschild generated by Hertz potentials and compare to the work of \cite{Pasq,PW-em,G-em}.

We will work with the mostly negative signature \((+,-,-,-)\) of the spacetime metric. We use the NP \cite{NP} and GHP \cite{GHP} formalisms throughout following the notation of \cite{GHP}. Otherwise, our conventions follow those of Wald \cite{Wald-book}. In situations where the meaning is clear, we will commonly omit writing the indices on a spacetime metric perturbation $\gamma_{ab}$, denoting it simply as $\gamma$.

\section{Schwarzschild background spacetime in the Carter frame}\label{sec:Sch-bg}

In any spacetime, consider a null frame \((l^a, n^a, m^a, \bar m^a)\) with \(l^a n_a = - m^a \bar m_a = 1\) and all other inner products vanishing. The metric can be written in terms of the tetrad as
\be
g_{ab} = 2l_{(a}n_{b)} - 2m_{(a}\bar m_{b)} 
\ee
We define the \emph{NP derivative operators} \((D, D', \delta, \delta')\) as \cite{NP}
\be\begin{split}
	D \defn l^a\nabla_a \eqsp D' \defn n^a\nabla_a \\
	\delta \defn m^a\nabla_a \eqsp \delta' \defn \bar m^a\nabla_a
\end{split}\ee

We recall the notion of \emph{GHP-weights} of tensor fields: Any tensor field \(\xi_{a\ldots}{}^{b\ldots}\) associated to the choice of null frame is said to have a GHP-weight \((p,q)\) if for any complex scalar \(\lambda\), under the transformations
\be\label{eq:GHP-transform}
	l^a \mapsto \lambda \bar\lambda l^a \eqsp n^a \mapsto (\lambda \bar\lambda)^{-1} n^a \eqsp m^a \mapsto \lambda (\bar\lambda)^{-1} m^a
\ee
the tensor \(\xi_{a\ldots}{}^{b\ldots}\) transforms as
\be
	\xi_{a\ldots}{}^{b\ldots} \mapsto \lambda^p \bar\lambda^q \xi_{a\ldots}{}^{b\ldots}
\ee
We will denote the GHP-weight of such tensors as \(\xi_{a\ldots}{}^{b\ldots} \wt (p,q)\). Note that 
\be
	g_{ab} \wt (0,0) \eqsp l^a \wt (1,1) \eqsp n^a \wt (-1,-1)\eqsp m^a \wt (1,-1) \eqsp \bar m^a \wt (-1,1)
\ee
For any GHP-weighted scalar \(\xi \wt (p,q)\) the \emph{GHP derivative operators} \((\thorn, \thorn',\eth,\eth')\) are defined by
\be\label{eq:GHP-d}\begin{split}
	\thorn\xi & \defn (D - p\epsilon - q \bar\epsilon)\xi \wt (p+1,q+1) \\
	\eth\xi & \defn (\delta - p\beta + q \bar\beta')\xi \wt (p+1,q-1) \\
	\thorn'\xi & \defn (D' + p\epsilon' + q \bar\epsilon')\xi \wt (p-1,q-1) \\
	\eth'\xi & \defn (\delta' + p\beta' - q \bar\beta)\xi \wt (p-1,q+1)
\end{split}\ee
where the various spin coefficients are as defined in \cite{GHP}.

In Schwarzschild coordinates the metric of the exterior of a Schwarzschild black hole of mass \(M\) is
\be\label{eq:Sch-metric}
	ds^2 = \frac{\Delta}{r^2} dt^2 - \frac{r^2}{\Delta}dr^2 - r^2 \lb( d\theta^2 + \sin^2\theta d\phi^2 \rb)
\ee
with \(\Delta \defn r(r - 2M)\). For the Schwarzschild metric \cref{eq:Sch-metric} we will use the Carter frame \cite{Carter-frame}
\be\label{eq:Carter-frame}\begin{split}
	l^a \equiv \tfrac{1}{\sqrt{2}r} \lb( \tfrac{r^2}{\sqrt{\Delta}}\partial_t + \sqrt{\Delta}\partial_r  \rb) \eqsp
	n^a \equiv \tfrac{1}{\sqrt{2}r} \lb( \tfrac{r^2}{\sqrt{\Delta}}\partial_t - \sqrt{\Delta}\partial_r  \rb) \eqsp 
	m^a \equiv \tfrac{1}{\sqrt{2}r} \lb( \partial_\theta + \tfrac{i}{\sin\theta}\partial_\phi \rb) 
\end{split}\ee
and the corresponding coframe
\be\label{eq:Carter-coframe}\begin{split}
	l_a \equiv \tfrac{1}{\sqrt{2}r} \lb( \sqrt{\Delta} dt - \tfrac{r^2}{\sqrt{\Delta}}dr \rb) \eqsp
	n_a \equiv \tfrac{1}{\sqrt{2}r} \lb( \sqrt{\Delta} dt + \tfrac{r^2}{\sqrt{\Delta}}dr \rb) \eqsp
	m_a \equiv - \tfrac{1}{\sqrt{2}} r \lb( d\theta + i\sin\theta d\phi \rb)
\end{split}\ee
The non-vanishing spin coefficients in the Carter frame are 
\be\label{eq:Carter-spin}\begin{split}
	\rho = -\rho' = -\frac{\sqrt{\Delta}}{\sqrt{2}r^2} \eqsp
	\beta = \beta' =  \frac{\cot\theta}{2\sqrt{2}r} \eqsp \epsilon = -\epsilon' = \frac{M}{2\sqrt{2\Delta}r}
\end{split}\ee
and the only non-vanishing curvature component is
\be\label{eq:Carter-curv}
	\Psi_2 \defn -C_{abcd}l^a m^b \bar m^c n^d = - \frac{M}{r^3} \wt (0,0)
\ee

The GHP commutators, Ricci and Bianchi identites on Schwarzschild give \cite{NP,GHP,Aks} 
\begin{subequations}\label{eq:Carter-ids1}\begin{align}
	[\thorn\eth - \eth\thorn]\xi = \bar\rho \eth \xi & \eqsp [\thorn\thorn' - \thorn'\thorn] \xi = - (p \Psi_2 + q \bar \Psi_2) \xi \label{eq:GHP-comm} \\
	\thorn \rho = \rho^2 \eqsp \thorn' \rho & = \rho \bar\rho' - \Psi_2 \eqsp \eth\rho = 0 \label{eq:Dspin} \\
     \delta \beta' + \delta' \beta & = \rho \rho' - \abs{\beta}^2 - \abs{\beta'}^2 - 2 \beta \beta' + \Psi_2      \label{eq:delta-beta1} \\
	\thorn \Psi_2 = 3 \rho \Psi_2 &\eqsp \thorn' \Psi_2 = 3 \rho' \Psi_2 \label{eq:DPsi2}
\end{align}\end{subequations}
where in \cref{eq:GHP-comm} \(\xi \wt (p,q)\).
In the Carter frame, \cref{eq:delta-beta1} simplifies to
\be
 \delta \beta = \tfrac{1}{2} (-\rho^2 + \Psi_2) - 2\beta^2
   \label{eq:delta-beta2}
\ee
For later convenience, we also define the derivative operators 
\be\begin{split}
	D_t \defn D + D' \eqsp D_r \defn D - D'
\end{split}\ee
which, in the Carter frame, act on a scalar \(\xi\) as
\be
	D_t \xi = \tfrac{\sqrt{2}r}{\sqrt{\Delta}} \partial_t \xi \eqsp D_r \xi = \tfrac{\sqrt{2\Delta}}{r} \partial_r \xi
\ee\\

Consider the Cauchy surface \(\Sigma\) defined by \(t=0\). The induced \emph{negative definite} metric on \(\Sigma\) is
\be
	h_{ab} = - r_a r_b - 2 m_{(a}\bar m_{b)}
\ee
where, in the Carter frame, 
\be
	r_a = \tfrac{1}{\sqrt{2}}(n_a - l_a) 
\ee
is the unit radial normal satisfying \(r_ar^a = -1\). The lapse function on \(\Sigma\) is
\be\label{eq:lapse}
	N = \frac{\sqrt{\Delta}}{r} = -\sqrt{2}r \rho
\ee
We denote by \(D_a\) the covariant derivative operator compatible with \(h_{ab}\). The induced metric on the spheres of constant \(r\) is
\be
	s_{ab} = - 2 m_{(a}\bar m_{b)}
\ee
and we denote the corresponding covariant derivative by \(\ms D_a\).

On \(\Sigma\) we have the following useful identities in the Carter frame
\begin{subequations}\label{eq:Carter-ids}\begin{align}
	D_ar^a = 2\sqrt{2}\rho & \eqsp D_b m^a = \ms D_b m^a = 2\beta m^a (m_b - \bar m_b) \label{eq:Dr-Dm} \\
	 D_a N = - r\Psi_2 r_a & \eqsp 4\epsilon \rho = \Psi_2 \label{eq:DN-epsilonrho}
\end{align}\end{subequations}
For any scalar \(\xi\) on \(\Sigma\) which is smooth on $\Sigma$ (including on the bifurcation surface, \(B\)) we have the following useful identities, which are obtained by integrating-by-parts and using \cref{eq:Carter-spin,eq:delta-beta2,eq:lapse,eq:Carter-ids}
\begin{subequations}\label{eq:IBP}\begin{align}
	\int_S (\delta + 2\beta) \xi = 0 \eqsp 2\int_S \beta \delta \xi & = - \int_S (-\rho^2 + \Psi_2)\xi \label{eq:IBP-sphere} \\
	\int_\Sigma N \rho D_r \xi & = -2 \int_\Sigma N (-\rho^2 + \Psi_2)\xi \label{eq:IBP2}
\end{align}\end{subequations}
Here \(S\) is any \(2\)-sphere of \(r=\text{constant}\), and in \cref{eq:IBP2} we have assumed that \(\xi = O(1/r^{(1+\epsilon)})\) near spatial infinity. Further, for any \emph{axisymmetric} scalar \(\xi\) we have
\be\label{eq:axisymm-id}
	\delta\xi = \delta'\xi
\ee

The Carter frame \cref{eq:Carter-frame} does not have a smooth limit to the past or future horizon, and, in particular, does not have a smooth limit to the bifurcation surface \(B = \partial \Sigma\). It is possible to rescale the Carter frame by a GHP-transformation \cref{eq:GHP-transform} with 
\be\label{eq:GHP-transform-smooth-frame}
	\lambda = \bar\lambda = \lb( \frac{\sqrt{\Delta}}{r} \rb)^{1/2}
\ee
so that it has a smooth limit to the future horizon (excluding $B$). Similarly, the inverse of this rescaling yields a frame that is smooth at the past horizon (excluding $B$). However, no frame that is Lie derived by the timelike Killing field $\partial_t$ can have a smooth limit to $B$, since a null vector at $B$ that is orthogonal to $B$ cannot be invariant under the isometries. Nevertheless, the rescaled null vector fields
\be\label{eq:smooth-null}
	\tilde l^a = \sqrt{\Delta} l^a \eqsp \tilde n^a = \sqrt{\Delta} n^a 
\ee
are smooth at $B$. This fact will be useful for determining the smoothness of quantities defined relative to the Carter frame.

\section{Hertz potentials for gravitational perturbations}\label{sec:hertz}

In the first subsection of this section, we will review the Hertz potential formalism for generating metric perturbations, following the formulation of \cite{Wald-potential}. In the second subsection we will address the issue of whether all nontrivial metric perturbations of Schwarzschild arise from a Hertz potential.

\subsection{Teukolsky equation, adjoints, and Hertz potentials}
\label{sec:Teu-adj-hertz}

Consider an algebraically special background spacetime solution to the vacuum Einstein equation and let \(\hat\gamma_{ab}\) be a metric perturbation. The linearized Einstein operator is given by
\be\label{eq:Ein-op}
	\mc E [\hat\gamma_{ab}] \defn - \nabla^2 \hat\gamma_{ab} - \nabla_a \nabla_b \hat\gamma + 2 \nabla^c \nabla_{(a}\hat\gamma_{b)c} + g_{ab} \lb( \nabla^c \nabla_c \hat\gamma - \nabla^c \nabla^d \hat\gamma_{cd} \rb)
\ee
Choose a null frame \((l^a, n^a, m^a, \bar m^a)\) with $l^a$ aligned with the repeated principal null direction of the background Weyl tensor. For the case of Petrov type-D spacetimes, such as Schwarzschild spacetime, we will also take $n^a$ to be aligned with the other repeated principal null direction. Let \(\hat\psi_0\) be the perturbation of the Weyl scalar \(\Psi_0 = - C_{abcd} l^a m^b l^c m^d \) corresponding to \(\hat\gamma_{ab}\), and define the operator
\be
	\mc T [\hat\gamma_{ab}] \defn \hat\psi_0 \wt (4,0)
\ee
The spin-\((2)\) Teukolsky operator is
\be
	\mc O[\hat\psi_0] \defn \lb[ (\thorn - 4\rho - \bar\rho)(\thorn' - \rho') - (\eth - 4\tau - \bar\tau')(\eth' - \tau') - 3 \Psi_2\rb]\hat\psi_0 \wt (4,0)
\ee

Teukolsky \cite{Teu,Teu-more} showed that if \(\hat\gamma_{ab}\) satisfies the linearized Einstein equation \(\mc E[\hat\gamma_{ab}] = 0\) then \(\hat\psi_0 = \mc T[\hat\gamma_{ab}]\) satisfies the ``decoupled'' spin-\((2)\) Teukolsky equation\footnote{Teukolsky \cite{Teu-more} assumed that the spacetime is Petrov type-D, but the same derivation works for any algebraically special vacuum spacetime.} \(\mc O[\hat\psi_0] = 0\). As noted in \cite{Wald-potential}, Teukolsky's derivation implies that there exists an operator \(\mc S\) such that the following operator identity holds
\be\label{eq:SEOT}
	\mc S \mc E = \mc O \mc T
\ee
The operator \(\mc S\) can be read off from the source terms in the \emph{inhomogenous} Teukolsky equation. We have (see Eqs.~2.12 and 2.13 of \cite{Teu-more} for the type-D case and Eq.~6.10 of \cite{Chrz} for the general algebraically special case)  
\be\begin{split}
	\mc S[T_{ab}] & \defn
 (\eth - 4\tau - \bar\tau') \lb[  - (\eth - \bar\tau')(l^al^b T_{ab}) + (\thorn - 2\bar\rho)(l^am^b T_{ab}) \rb] \\
		&\qquad + (\thorn - 4\rho - \bar\rho) \lb[ \bar\sigma' (l^al^bT_{ab}) + (\eth - 2\bar\tau')(l^am^bT_{ab}) - (\thorn - \bar\rho)(m^am^bT_{ab}) \rb] \wt (4,0)
\end{split}\ee

To state the main result from \cite{Wald-potential} we introduce the following notion of adjoints. Consider any linear differential operator \(\mc L\) mapping \(n\)-index (possibly complex) tensor fields \(\xi_{a_1 \ldots a_n}\) to \(m\)-index (possibly complex) tensor fields \((\mc L[\xi])_{a_1 \ldots a_m}\). We define the (real\footnote{Note that the definition of adjoint does not include any complex conjugation.}) adjoint, \(\adj{\mc L}\), of \(\mc L\) to be the linear differential operator mapping \(m\)-index tensor fields \(\chi^{a_1 \ldots a_m}\) to \(n\)-index tensor fields \((\adj{\mc L} [\chi])^{a_1 \ldots a_n}\) that satisfies
\be\label{eq:adj-defn}
	\chi^{a_1 \ldots a_m} (\mc L [\xi])_{a_1 \ldots a_m} - (\adj{\mc L} [\chi])^{a_1 \ldots a_n} \xi_{a_1 \ldots a_n} = \nabla_a v^a (\chi, \xi)
\ee
where \(v^a\) is a vector field that is a bilinear in \(\chi\) and \(\xi\) and is locally constructed from these quantities and their derivatives. If $\mc L$ is defined on all tensor fields $\xi_{a_1 \ldots a_m}$ of its tensorial type (i.e., if there are no differential or other relations that $\xi_{a_1 \ldots a_m}$ must satisfy), then $\adj{\mc L}$ is uniquely defined by \cref{eq:adj-defn}, but \(v^a\) is defined only up to addition of a term of the form $\nabla_b V^{ab}(\chi, \xi)$ with $V^{ab} = - V^{ba}$. 
We call a operator \(\mc L\) self-adjoint if \(\mc L = \adj{\mc L}\), using the background spacetime metric to raise or lower indices if needed.

Taking the adjoint of \cref{eq:SEOT} we get the operator identity
\be\label{eq:TOES}
	\adj{\mc T} \adj{\mc O} = \adj{\mc E} \adj{\mc S} 
\ee
Here, the adjoints $\adj{\mc E}$, $\adj{\mc O}$ and $\adj{\mc T}$ are uniquely defined by \cref{eq:adj-defn}. However, since $\mc S$ in \cref{eq:SEOT} acts only on divergence-free symmetric tensor fields (since $\mc E [\gamma_{ab}]$ is divergence-free for any $\gamma_{ab}$), $\adj{\mc S}$ is ambiguous up to addition of a term of the form $\nabla_{(a} \eta_{b)}$.

The linearized Einstein operator \cref{eq:Ein-op} is self-adjoint, \(\adj{\mc E} = \mc E\). From \cref{eq:TOES} we see that, if \(\psi \wt (-4,0)\) is a solution to the equation \(\adj{\mc O}[\psi] = 0\), then \(\gamma^{ab} \defn \adj{\mc S}[\psi] \) is a solution of the linearized Einstein equation \(\mc E[\gamma_{ab}] = 0\). Note that $\gamma_{ab}$ is complex. We call such a \(\gamma_{ab}\) the \emph{complex metric perturbation generated by the Hertz potential} \(\psi\), whereas we will refer to $\Re \adj{\mc S}[\psi] $ as the \emph{real metric perturbation generated by} \(\psi\).

Explicitly computing the adjoint of $\mc O$, we obtain
\be\label{eq:Teu-op}
	\adj{\mc O}[\psi] = \lb[ (\thorn' - \bar\rho')(\thorn + 3\rho ) - (\eth' - \bar\tau )(\eth + 3\tau ) - 3 \Psi_2 \rb]\psi = 0
\ee
For the case of type-D spacetimes, it can be seen that see that $\adj{\mc O}$ is precisely the spin-\((-2)\) Teukolsky operator.
The complex metric perturbation generated by \(\psi\) is
\be\label{eq:metric-Hertz}\begin{split}
	\gamma^{ab} \defn \adj{\mc S}[\psi] & = \lb[ -l^al^b ( \eth - \tau ) + l^{(a}m^{b)} ( \thorn - \rho + \bar\rho ) \rb]( \eth + 3\tau )\psi \\
	&\quad + \lb[ - l^al^b \bar\sigma' + l^{(a}m^{b)} ( \eth - \tau + \bar\tau' ) - m^a m^b ( \thorn - \rho ) \rb]( \thorn + 3\rho )\psi \wt (0,0)
\end{split}\ee
where we have used the ambiguity in $\adj{\mc S}$ to define it so that $\gamma_{ab}$ satisfies the ``ingoing radiation'' gauge conditions
\be\label{eq:ingoing-gauge}
	\gamma_{ab} l^b = 0 \eqsp g^{ab}\gamma_{ab} = 0
\ee

On a Schwarzschild background the Teukolsky equation \cref{eq:Teu-op} takes the form\footnote{On Schwarzschild, the Teukolsky equation \cref{eq:Teu-op-Sch} is the decoupled equation for the Weyl curvature component first found by Bardeen and Press \cite{BP}.}
\be\label{eq:Teu-op-Sch}
	\adj{\mc O}[\psi] = \lb[ (\thorn' - \bar\rho')(\thorn + 3\rho ) - \eth'\eth - 3 \Psi_2\rb]\psi = 0
\ee
and the complex metric perturbation generated by \(\psi\) \cref{eq:metric-Hertz} is
\begin{subequations}\label{eq:metric-Hertz-Sch}\begin{align}
	\gamma_{ab} & = -l_al_b U + l_{(a}m_{b)} V - m_a m_b W \quad\text{with} \label{eq:metrix-UVW} \\[1.5ex]
	\begin{split}\label{eq:U-V-W-defn}
	U & \defn \eth^2 \psi \wt (-2,-2) \\
	V & \defn \lb[ \thorn \eth +  \eth ( \thorn + 3\rho ) \rb]\psi \wt (-2,0)\\
	W & \defn ( \thorn - \rho ) ( \thorn + 3\rho )\psi \wt (-2,2) \\
	\end{split}
\end{align}\end{subequations}

Writing $l^a = {\tilde l}^a /\sqrt{\Delta}$, we see from \cref{eq:smooth-null} that $\gamma_{ab}$ will be smooth at $B$ provided that
\be\label{eq:U-V-W-smooth}
	U = \Delta \tilde U \eqsp V = \sqrt{\Delta} \tilde V \eqsp W = \tilde W \eqsp 
\ee
where \(\tilde U, \tilde V, \tilde W\) are smooth at $B$. This will hold, in turn, provided that 
\be\label{eq:hertz-smooth}
\psi = \Delta \tilde\psi
\ee
where $\tilde \psi$ is smooth at $B$.

\subsection{Generation of metric perturbations by Hertz potentials}
\label{sec:all-hertz}

We conjecture that --- modulo trivial perturbations that are a linear combination of a pure-gauge perturbation and a perturbation to some Kerr spacetime --- any real, smooth metric perturbation of Schwarzschild can be obtained as the real part of a metric perturbation generated by a smooth (with the rescaling indicated at the end of \cref{sec:Sch-bg}) Hertz potential. If true, then one could analyze stability of Schwarzschild without any further assumptions by restricting consideration to perturbations generated by a Hertz potential. However, we have not been able to show this.\footnote{It is possible to directly show that all real frequency metric perturbations can be generated by a Hertz potential \cite{Wald-sd}. However, this result clearly is not adequate for carrying out a stability analysis.} In this subsection, we will argue that any (real) metric perturbation with initial data that is smooth and of compact support can be approximated arbitrarily well in an $L^2$ norm on initial data by a real metric perturbation generated by a smooth Hertz potential with initial data of compact support plus a gauge transformation. Our arguments are a precise version of the heuristic discussion given in \cite{Wald-potential-RN}. Our arguments fail to be a proof only in that we will not attempt to prove that certain singular solutions that cannot be generated by a Hertz potential fail to have representatives in our $L^2$ space.

Our arguments will be based on a relationship between the symplectic product of solutions to the linearized Einstein equation and a corresponding bilinear product of solutions to the spin-$(2)$ and spin-$(-2)$ Teukolsky equations. From our general definition of adjoints, \cref{eq:adj-defn}, we know that for any two metric perturbations $\gamma_{ab}$ and $\gamma'_{ab}$, there exists a $w^a$ locally constructed from $\gamma_{ab}$, $\gamma'_{ab}$ and their derivatives such that
\be\label{eq:adj-currents1}
\gamma{}^{ab} (\mc E[\gamma'])_{ab} - (\adj{\mc E}[\gamma])^{ab} \gamma'_{ab}  = \nabla_a w^a(g; \gamma, \gamma')
\ee
Since $\mc E$ is self-adjoint, we may drop the $\dagger$ on $\adj{\mc E}$. In fact, it is easily seen from the Lagrangian formulation of general relativity \cite{IW} that $w^a$ is simply the symplectic current associated with the Lagrangian and is given by the explicit expression
\be\label{eq:w-defn}
w^a  = P^{abcdef} \lb( \gamma'_{bc} \nabla_d \gamma_{ef} - \gamma_{bc} \nabla_d \gamma'_{ef} \rb) 
\ee
where
\be
	P^{abcdef} = g^{ae}g^{fb}g^{cd} - \tfrac{1}{2} g^{ad}g^{be}g^{fc} - \tfrac{1}{2} g^{ab}g^{cd}g^{ef} - \tfrac{1}{2} g^{bc}g^{ae}g^{fd} + \tfrac{1}{2} g^{bc}g^{ad}g^{ef} \, .
\ee
The symplectic product of $\gamma_{ab}$ and $\gamma'_{ab}$ is obtained by integrating $w^a$ over a Cauchy surface $\Sigma$ for the exterior of the black hole
\be \label{Omega}
 \Omega(\gamma, \gamma') \defn \int_\Sigma u_a w^a(\gamma, \gamma')
\ee
where $u^a$ denotes the unit normal to $\Sigma$. The symplectic form \(\Omega\) is non-degenerate on the space of linearized solutions to Einstein's equation modulo gauge. In terms of the linearized initial data $(p_{ab}, q_{ab})$, we have
\be \label{Omega2}
 \Omega(\gamma, \gamma') = \int_\Sigma (p^{ab} q'_{ab} - q_{ab} p'^{ab})
\ee
It follows immediately from \cref{eq:adj-currents1} that $\nabla_a w^a = 0$ when $\gamma_{ab}$ and $\gamma'_{ab}$ satisfy the linearized Einstein equation, so  \(\Omega\) is conserved, i.e., independent of the choice of Cauchy surface \(\Sigma\).

Similarly, for the Teukolsky operator \(\mc O\), we have
\be\label{eq:adj-currents2}
		\psi \mc O[\psi'] - \adj{\mc O}[\psi] \psi' = \nabla_a \pi^a(g; \psi, \psi')
\ee
where \(\psi' \wt (4,0)\) and \(\psi \wt (-4,0)\) and $\pi^a$ is locally constructed out of $g$, $\psi$, and $\psi'$. By explicit computation, we obtain 
\be\label{eq:pi-exp}
	\pi^a(\psi,\psi') = \psi (\thorn' - \rho') \psi' l^a - (\thorn + 3\rho) \psi \psi' n^a - \psi (\eth'-\bar\tau')\psi' m^a + (\eth + 3\tau) \psi \psi' \bar m^a
\ee
We define
\be \label{Pi}
     \Pi(\psi, \psi') \defn \int_\Sigma u_a \pi^a(\psi, \psi')
\ee
Evaluating \(\Pi(\psi,\psi')\) on the \(t=0\) Cauchy surface in Schwarzschild, using \(u_a = \tfrac{1}{\sqrt{2}} (n_a + l_a)\), we get
\be \label{Pi-non-degen}
	\Pi(\psi, \psi') = - \tfrac{1}{\sqrt{2}} \int_\Sigma \lb( \eta \psi' - \psi \eta' \rb)
\ee
where
\be \label{eta}
\eta \defn (\thorn+3\rho)\psi \wt (-3,1) \eqsp \eta' \defn (\thorn' - \rho') \psi' \wt (3,-1)
\ee
Since, \((\psi, \eta) , (\psi', \eta')\) are free initial data for the Teukolsky equations, \(\Pi\) is manifestly non-degenerate. It follows immediately from \cref{eq:adj-currents2} that, for solutions to
the Teukolsky equations \(\adj{\mc O} [\psi] = 0\), \(\mc O[\psi'] = 0\), we have $\nabla_a \pi^a = 0$, so \(\Pi\) is independent of the choice of Cauchy surface \(\Sigma\).

Following \cite{Wald-potential-RN} we obtain the following relation between \(\Omega\) and \(\Pi\):
\begin{prop}\label{prop:symp-id}
Let \(\psi\) be any smooth solution to the spin-(\(-2\)) Teukolsky equation \(\adj{\mc O}[\psi] = 0 \) with initial data of compact support,\footnote{The Cauchy surface $\Sigma$ for the exterior region does not include the bifurcation surface $B=\partial \Sigma$, so ``compact support initial data'' here and elsewhere requires the initial data on $\Sigma$ to vanish in a neighborhood of $B$.} and let \(\gamma_{ab}\) be a smooth, real perturbation (not necessarily of compact support) solving the linearised Einstein equation \(\mc E[\gamma] = 0\). Then we have
\be\label{eq:symp-id}
    \Omega(\Re(\adj{\mc S}[\psi]), \gamma)  = \Re \lb[ \Pi(\psi, \mc T[\gamma]) \rb]
\ee
where $\mc S$ and $\mc T$ are as in \cref{eq:SEOT}.
\begin{proof}
It is convenient to extend the definition of $w^a$ and $\Omega$ to complex metric perturbations by complex linearity\footnote{We continue to define the adjoint by \cref{eq:adj-defn} (with no complex conjugations) when considering complex perturbations, so ``taking the adjoint'' remains a linear (rather than antilinear) operation.} in each variable. Let \(\gamma'_{ab}\) be an arbitrary smooth, complex metric perturbation that does not necessarily satisfy the linearized Einstein equation. Consider the following equations that follow immediately from the definition, \cref{eq:adj-defn}, of adjoint operators
\begin{subequations}\begin{align}
    \psi (\mc S \mc E[\gamma'])  - \adj{\mc S}[\psi] \mc E[\gamma'] & = \nabla_a s^a(\psi, \mc E[\gamma']) \\
    \adj{\mc S}[\psi] \mc E[\gamma'] - (\adj{\mc E} \adj{\mc S}[\psi]) \gamma' & = \nabla_a w^a (\adj{\mc S}[\psi], \gamma') \\
    \psi (\mc O \mc T[\gamma']) - \adj{\mc O}[\psi] \mc T[\gamma'] & = \nabla_a \pi^a(\psi, \mc T[\gamma'])
\end{align}\end{subequations}
We add the first two of these equations and subtract the third, using $\mc S \mc E = \mc O \mc T$ (see \cref{eq:SEOT}), $\adj{\mc O}[\psi] = 0$, and $\adj{\mc E} \adj{\mc S}[\psi] = \adj{\mc T}\adj{\mc O}[\psi] = 0$. We obtain
\be
\nabla_a \left(s^a + w^a - \pi^a \right) = 0 \, .
\ee
Since this holds for arbitrary $\gamma'_{ab}$, by the results of \cite{W-closed}, there exists an $H^{ab} = - H^{ba}$ locally constructed out of $g_{ab}$, $\psi$, and $\gamma'_{ab}$ such that\footnote{The dualized form of this result, in terms of differential forms, is that if $d(*s + *w + * \pi) = 0$ then $*s + *w + * \pi = d*H$. No assumption about topology is used to prove this result.}
\be
s^a + w^a - \pi^a = \nabla_b H^{ab}
\ee
Integrating over $\Sigma$, we obtain
\be \label{symid}
\int_\Sigma u_a \left(s^a + w^a - \pi^a \right) = \int_\Sigma u_a \nabla_b H^{ab} = \int_{\partial \Sigma} u_a r_b H^{ab} = 0 
\ee
where the last equality follows from the fact that $H^{ab}$ vanishes on the boundary because $\psi$ is of compact support on $\Sigma$. 
\cref{symid} holds for arbitrary $\gamma'_{ab}$. We now specialize to the case 
$\gamma'_{ab} = \gamma_{ab}$, where $\gamma_{ab}$ is a real solution to the linearized Einstein equation. Then $\mc E [\gamma] = 0$, so $s^a (\psi, \mc E[\gamma]) = 0$. Using the above definitions \cref{Omega,Pi} of $\Omega$ and $\Pi$, we obtain
\be\label{eq:Omega-Pi}
    \Omega(\adj{\mc S}[\psi], \gamma)  = \Pi(\psi, \mc T[\gamma])
\ee
Taking the real part of this equation, we obtain the desired result.
\end{proof}
\end{prop}

For the analysis of positivity of canonical energy, we wish to consider perturbations that lie in the space\footnote{In \cite{PW}, a further space $\ms V^\infty$ was considered which uniquely fixed the gauge in $\ms V_c^\infty$. However, it would not be convenient to impose these gauge conditions here, since $\Re(\adj{\mc S}[\psi])$ will not satisfy these gauge conditions.} $\ms V_c^\infty$ constructed in Sec.~3  of \cite{PW}. This space is defined as follows. We start with the real $L^2$ space $\ms V_0$ of all (unconstrained, real) initial data, with inner product
\be \label{L2}
\big\langle (p'_{ab}, q'_{ab}) \big | (p_{ab}, q_{ab}) \big\rangle \defn \int_{\Sigma} (p'^{ab} p_{ab} + q'^{ab} q_{ab})
\ee
On $\ms V_0$, $\Omega: \ms V_0 \times \ms V_0 \to \bb R$ is a bounded bilinear form, so it corresponds to a bounded linear map $\tilde{\Omega}: \ms V_0 \to \ms V_0$. It is easily seen from \cref{Omega2} that this map is anti-self-adjoint and orthogonal i.e., $\adj{\tilde{\Omega}} = -{\tilde{\Omega}}$ and $\adj{\tilde{\Omega}}\tilde{\Omega} = I$, where here the adjoint \(\dagger\) is in the \(L^2\) inner product,. Let $\ms W_c$ be the subspace of pure gauge metric perturbations \(\Lie_\xi g_{ab}\) generated by smooth vector fields \(\xi^a\) that become an asymptotic translation or a rotation at infinity and whose projection onto $B$ vanishes \cite{PW}. Let $\ms V_c$ be the subspace of $\ms V_0$ that is symplectically orthogonal to $\ms W_c$. Elements of $\ms V_c$ (weakly) satisfy the constraints, as well as\footnote{Here and in \cref{eq:conditions}, ``\(\delta\)'' represents perturbed quantities and should not be confused with the NP derivative operator \(\delta\).} $\delta M = \delta J = \delta P_i = 0$ and the following boundary conditions\footnote{In fact, as shown in \cite{Sorce-Wald}, the condition $\delta \varepsilon = 0$ is not needed. However, no harm is done by imposing this condition, since it is merely a gauge condition when $\delta M = \delta J = 0$.} at $B$ \cite{HW-stab, PW}
\be\label{eq:conditions}
	\delta \varepsilon = \delta \vartheta_+ = \delta \vartheta_- = 0
\ee
Here \(\delta\varepsilon\) is the perturbed area element of the bifurcation surface \(B\), and \(\delta\vartheta_{\pm}\) are the perturbed outgoing and ingoing expansions of \(B\). Only \(\delta M = \delta J = 0\) are physical restrictions, as the conditions \cref{eq:conditions} can always be achieved by a choice of gauge. Finally, $\ms V_c^\infty$ is obtained by intersecting $\ms V_c$ with appropriate weighted Sobolev spaces, so that $\ms V_c^\infty$ consists of smooth solutions to the constraints that satisfy the boundary conditions \cref{eq:conditions} at $B$ and fall off as \(p_{ab}, q_{ab} = o(1/r^{\nfrac{3}{2}})\) at spatial infinity (with all spatial derivatives falling off faster by corresponding powers of \(r\)). Note that the fall off in $q_{ab}$ is faster than usually assumed but does not impose any undesireable restrictions in our analysis since we are only interested in perturbations with \(\delta M = 0\). The space $\ms V_c^\infty$ is dense in $\ms V_c$ \cite{PW}. We refer the reader to \cite{PW} for the details of the construction and to \cite{HW-stab,PW} for an explanation as to why \cref{eq:conditions} and the conditions $\delta M = \delta J = \delta P_i = 0$ are imposed.

Let $\Omega_c$ denote the restriction of $\Omega$ to the real Hilbert space $\ms V_c$. It follows from the construction of $\ms V_c$ that $\Omega_c$ is degenerate precisely on elements that lie in $\ms V_c \inter \bar{\ms W}_c$ where the bar denotes the closure of the subspace. These elements are (limits of) pure gauge/symmetry perturbations. Let $\tilde{\Omega}_c : \ms V_c \to \ms V_c$ be the linear map corresponding to $\Omega_c$. Then $\tilde{\Omega}_c$ is a bounded, anti-self-adjoint map, although it is not orthogonal.

Let \(\psi\) be a smooth solution to the spin-(\(-2\)) Teukolsky equation \(\adj{\mc O}[\psi] = 0 \) with initial data of compact support. Then $\Re(\adj{\mc S}[\psi]) \in \ms V_c^\infty \subset \ms V_c$. Let $\ms Y$ be the subspace of $\ms V_c$ generated by such perturbations. Let $\ms X \subset \ms V_c$ denote the subspace of all smooth solutions to the linearized Einstein equation with initial data of compact support. 

We now will argue that given any $\gamma \in \ms X$ and given any $\epsilon > 0$, there exists a $\gamma_\psi \in \ms Y$ and a gauge perturbation $\gamma_\xi \in \ms V_c \inter \bar{\ms W}_c$ such that $||\gamma - \gamma_\psi - \gamma_\xi|| < \epsilon$. This would show that every smooth solution with initial data of compact support is well approximated in the $L^2$ norm \cref{L2} by a solution generated by a smooth Hertz potential with initial data of compact support plus a gauge transformation. This is equivalent to showing that
\be \label{XU}
\bar{\ms X} \subset \bar{\ms Y} + \left( \ms V_c \inter \bar{\ms W}_c \right) \, .
\ee
Since $\tilde{\Omega}_c$ is bounded and its kernel is precisely $\ms V_c \inter \bar{\ms W}_c$, this, in turn, is equivalent to showing that 
\be
\tilde{\Omega}\lb[\bar{\ms X}\rb] \subset \tilde{\Omega}\lb[\bar{\ms Y}\rb]  \, .
\ee
Finally, this, in turn, is equivalent to\footnote{Note that since $\ms Y \subset \ms X$, this and the previous set inclusion can hold only if equality holds.}
\be \label{symcom}
\left(\tilde{\Omega}[\bar{\ms X}] \right)^\perp  = \left(\tilde{\Omega}[\ms X] \right)^\perp \supset \left(\tilde{\Omega}[\bar{\ms Y}] \right)^\perp = \left(\tilde{\Omega}[\ms Y] \right)^\perp  \, .
\ee
where $\perp$ denotes the orthogonal complement. 

Now, we have $\gamma \in \left(\tilde{\Omega}[\ms Y] \right)^{\perp}$ if and only if for all smooth solutions $\psi$ to the spin-(\(-2\)) Teukolsky equation \(\adj{\mc O}[\psi] = 0 \) with initial data of compact support, we have
\be \label{symcom2}
\Omega\lb( \Re(\adj{\mc S}[\psi], \gamma) \rb) = 0
\ee
We will now argue (but not prove) that if $\gamma$ satisfies this relation, then $\gamma$ must be pure gauge. However, all gauge transformations are also in the symplectic complement of $\ms X$, so this would establish \cref{symcom}. Thus, we will prove \cref{XU} if we can show that any $\gamma$ that satisfies \cref{symcom2} must be pure gauge.

\cref{prop:symp-id} and the nondegenerate form \cref{Pi-non-degen} of $\Pi$ immediately imply that any $\gamma$ that satisfies \cref{symcom2} must (weakly) satisfy\footnote{\cref{prop:symp-id} was proven for smooth $\gamma$, whereas, a priori, the $\gamma$ appearing in \cref{symcom2} is only known to be in $\ms V_c$ (and, thus, locally in $L^1$ and hence a distribution). However, any $\gamma \in \ms V_c$ can be approximated in the $L^2$ norm by smooth elements of $\ms V_c$, and \cref{T0} then follows immediately.}
\be \label{T0}
\mc T[\gamma] = 0 \, ,
\ee
i.e., the Teukolsky variable $\psi_0$ obtained from $\gamma$ must vanish, so $\gamma$ corresponds to an algebraically special perturbation. By the Starobinski-Teukolsky identities (see \cite{AAB} with their \(\kappa_1 = -r/3\)), we have
\begin{subequations}\label{eq:STI}\begin{align}
	\eth'^4 (r^4 \psi_0) & = \thorn^4 (r^4 \psi_4) + 3M \Lie_t \bar\psi_0 \label{eq:STI1} \\
	\thorn'^4 (r^4 \psi_0) & = \eth^4 (r^4 \psi_4) + 3M \Lie_t \bar\psi_4 \label{eq:STI2}
\end{align}\end{subequations} 
Setting $\psi_0 = 0$ and writing \(\psi = r^4 \psi_4\), \cref{eq:STI1,eq:Teu-op} for \(\psi\) are 
\begin{subequations}\label{eq:eqns}\begin{align}
    \lb[ \lb(\partial_u - \tfrac{\Delta}{2r^2} \partial_r - \tfrac{r-4M}{2r^2} \rb) \lb(\partial_r - \tfrac{3}{r} \rb) - \eth'\eth - 3 \Psi_2 \rb] \lb( \tfrac{\Delta}{2r^2} \psi \rb) & = 0 \label{eq:eqns-Teu}\\
   \partial_r^4 \lb( \tfrac{\Delta}{2r^2} \psi \rb) & = 0 \label{eq:eqns-STI}
\end{align}\end{subequations}
where we have written these equations using \emph{outgoing Eddington-Finkelstein coordinates}, with \((u = t - r^*, r, \theta, \phi)\) with $r^*$ defined by $dr^*/dr = r^2/\Delta$. A priori, these equations for $\psi$ are known only to hold distributionally (where we now view the solution $\psi$ as a distribution on spacetime rather than a distribution on initial data). However, we can effectively expand $\tfrac{\Delta}{2r^2} \psi$ in spin-\((-2)\)-weighted spherical harmonic functions \(\mc Y^{(-2)}_{\ell, m}(\theta, \phi)\) with \(\ell \geq 2\) (see Sec.~4.15 \cite{PR1}) by smearing it with test functions of the form $f(u,r) \mc Y^{(-2)}_{\ell, m}$, and considering the resulting distribution $\psi_{\ell, m}$ on $f(u,r)$. The general solution to \cref{eq:eqns-STI} is a cubic polynomial in \(r\)
\be
    \psi_{\ell, m} = \sum\limits_{k=0}^{3} \alpha_k(u) r^k 
\ee
We can then substitute this into \cref{eq:eqns-Teu}, replacing $\eth'\eth$ by $L/r^2$ with $L = -\tfrac{1}{2}(\ell+2)(\ell - 1)$. We obtain the general solution
\be \label{algspec}
    \alpha_0 = C_0 e^{\omega u} \eqsp \alpha_1 = - \tfrac{L}{M} \alpha_0 \eqsp \alpha_2 = \tfrac{L^2}{3M^2} C_0 e^{\omega u} + C_2 e^{-\omega u} \eqsp \alpha_3 = - \tfrac{1}{L} \partial_u \alpha_2 
\ee
where \(\omega = \frac{L(L-1)}{3 M} = \frac{(\ell-1)\ell(\ell+1)(\ell+2)}{12M} > 0\) and \(C_0, C_2\) are constants. 

It can be seen by inspection of \cref{algspec} that if $C_2 \neq 0$, then $\psi$ blows up exponentially near spatial infinity ($r \to \infty$ with $t$ fixed, so $u \to - \infty$). On the other hand, if $C_0 \neq 0$, then $\psi$ is singular as one approaches $B$ (i.e., $r \to 2M$ with $t$ fixed, so $u \to + \infty$), as originally found by Couch and Newman \cite{CN}. Indeed, for $\ell = 2$, $\psi$ blows up as $(r-2M)^{-4}$, and the blow-up is faster for higher $\ell$. It seems clear that the initial data for any metric perturbation $\gamma$ corresponding to a nonvanishing solution to \cref{algspec} must fail to be square integrable in any gauge. However, we have not attempted to prove this.

Assuming that this is the case, we have found that any $\gamma$ for which \cref{symcom2} holds must have both Teukolsky variables vanish, $\psi_0 = \psi_4 = 0$. 
However, all perturbations with $\psi_0 = \psi_4 = 0$ were obtained in \cite{Wald-Kerr-pert}, where it was found that --- up to gauge --- the general solution is a linear combination of perturbations of the Kerr parameters, the NUT parameter, and the C-metric parameter (4 parameters total). The NUT and C-metric perturbations are singular. We believe that there are no metric representatives of these perturbations that have square integrable initial data, but we have not attempted to prove this. The Kerr perturbations are excluded by our construction of $\ms V_c$. Thus, we conclude that the only solutions $\gamma \in \ms V_c$ to \cref{symcom2} are pure gauge perturbations. This implies that every smooth solution with initial data of compact support can be approximated arbitrarily well in the $L^2$ norm on initial data by a solution generated by a smooth Hertz potential of compact support plus a gauge transformation.

\begin{remark}
\cref{prop:symp-id} holds for perturbations of any algebraically special vacuum spacetime (although $\adj{\mc O}[\psi] = 0$ would not have the interpretation of being a Teukolsky equation if the spacetime is not type-D). It is likely that a similar analysis of perturbations with $\psi_0 = 0$ could be given for Kerr (see \cite{Wald-Kerr-pert}). The analysis of perturbations with $\psi_0 = \psi_4 = 0$ applies to Kerr \cite{Wald-Kerr-pert}.
\end{remark}

\section{Positivity of the canonical energy for metric perturbations obtained from a hertz potential}
\label{sec:positive-can-energy}

The \emph{canonical energy} is a bilinear form on metric perturbations of a stationary black hole, defined as \cite{HW-stab}
\be\label{eq:can-energy-defn}
	\ms E(\gamma_1,\gamma_2) \defn \Omega(\gamma_1, \Lie_t\gamma_2)
\ee
where $\Omega$ is the symplectic product \cref{Omega}.
The canonical energy is symmetric, i.e. \(\ms E(\gamma_1,\gamma_2) = \ms E(\gamma_2,\gamma_1 )\) (see Prop.~2 \cite{HW-stab}), and conserved (as follows immediately from the conservation of the symplectic form). As explained in detail in \cite{HW-stab,PW}, we are interested in the positivity properties of $\ms E$ when acting on the space $\ms V_c^\infty \subset \ms V_c$ of smooth elements of $\ms V_c$ (defined in the previous section). When restricted to the space $\ms V_c^\infty$, \(\ms E\) is gauge-invariant and is degenerate precisely on perturbations to other stationary black holes \cite{HW-stab,PW}. Its positivity on $\ms V_c$ implies mode stability, whereas its failure to be positive on this space implies instability \cite{HW-stab,PW}.

We would like to show that $\ms E$ is positive on the space $\ms V_c^\infty$. However, as already discussed in the Introduction, we have not succeeded in directly showing the positivity of expression \cref{eq:PE-defn} for the ``potential energy.'' Since the main difficulty appears to arise from the fact that $q_{ab}$ is not a free variable but must satisfy the constraint \cref{eq:lin-Ham-constraint}, a promising strategy is to consider the canonical energy of perturbations generated by a Hertz potential $\psi$, since the canonical energy would then be expressed purely in terms of the unconstrained variable $\psi$. Thus, we consider metric perturbations of the form $\gamma = \Re(\adj{\mc S}[\psi])$. We will now show that the canonical energy of such perturbations is positive.

Although our interest is in real perturbations of the above form, it is convenient to allow complex metric perturbations and extend the definition of canonical energy \(\ms E(\gamma_1,\gamma_2)\) to complex perturbations by taking it to be antilinear in its first variable and linear in its second. This makes $\ms E$ a \emph{Hermitian form} i.e., \(\ms E(\gamma_1,\gamma_2) = \bar{\ms E(\gamma_2,\gamma_1)}\).
Explicitly, for complex initial data $(p_{ab}, q_{ab})$, the kinetic and potential energies are given by\footnote{Again, we remind the reader that our spatial metric is negative definite.}
\begin{subequations}\begin{align}
	\ms K & = \tfrac{1}{16\pi} \int_\Sigma 2 N~ \lb[ \bar p_{ab} p^{ab} - \tfrac{1}{2}\bar p p  \rb] \label{eq:complex-KE-defn}\\
	\ms U & = -\tfrac{1}{16\pi} \int_\Sigma N \lb[ \tfrac{1}{2} D_c\bar q_{ab} D^cq^{ab} - D_c\bar q_{ab} D^a q^{cb} - \tfrac{3}{2} D_c\bar q D^cq + D^c \bar q D^aq_{ac} + D^a\bar q_{ac} D^c q \rb] \label{eq:complex-PE-defn}
\end{align}\end{subequations}

We now compute the canonical energy of the complex metric perturbation $\gamma = \adj{\mc S}[\psi]$ in terms of the unconstrained initial data $(\psi, \eta)\vert_\Sigma$ for $\psi$, where $\eta$ was defined in \cref{eta}. We will show that the canonical energy of $\adj{\mc S}[\psi]$ is positive. We will then show that this implies that the canonical energy of $\Re(\adj{\mc S}[\psi])$ also is positive. 

To evaluate the canonical energy in terms of the initial data $(\psi, \eta)\vert_\Sigma$ for the Hertz potential we will eliminate any time derivatives of these quantities using the Teukolsky equation. Many of the resulting expressions will not be invariant under the GHP-transformations (\cref{eq:GHP-transform}) and will only hold in the Carter frame. Nevertheless, it is still convenient to use GHP derivatives on various properly GHP weighted quantities that occur in the resulting equations, as this simplifies the notation. Thus, some of our equations below will contain a mix of GHP and NP derivatives and, in general, they will be valid only in the Carter frame. 

From \cref{eq:metric-Hertz-Sch}, we see that the (complex) perturbed spatial metric $q_{ab}$ on $\Sigma$ generated by this perturbation, in the Carter frame, is given by
\be\label{eq:q-hertz}
	q_{ab} = -\tfrac{1}{2}r_ar_b U - \tfrac{1}{\sqrt{2}} r_{(a}m_{b)} V - m_a m_b W 
\ee
where
\begin{subequations}\label{eq:U-V-W-alt}\begin{align}
	U & = \eth^2\psi \\
	V & = 2 \eth \eta - 2\rho \eth \psi \\
	\label{eq:W-used-Teu1}
	W  & = (\thorn - \rho)\eta 
\end{align}\end{subequations}
Using the Teukolsky equation in the Carter frame,
\be\label{eq:Teu-initial}\begin{split}
	\thorn'\eta & = -\rho \eta + (\eth'\eth + 3 \Psi_2 )\psi
\end{split}\ee
we can eliminate the time derivative of $\eta$ and rewrite \cref{eq:W-used-Teu1} as
\be
W = (D_r - 2\rho)\eta + ( \eth'\eth + 3 \Psi_2)\psi
\label{eq:W-used-Teu2}
\ee
Note that \cref{eq:Teu-initial} can also be written as
\be\label{eq:Teu-W}
	\thorn'\eta =  W - (D_r - \rho)\eta
\ee

The (complex) perturbed ADM momentum \(p_{ab}\) on $\Sigma$ can be found using the linearized ADM evolution equation (see Eq.~5.4 \cite{PW})
\be
	\Lie_t q_{ab} = 2N\lb( p_{ab} - \tfrac{1}{2}p h_{ab} \rb) + 2 D_{(a}\lb( h_{b)}{}^c t^d \gamma_{cd} \rb)
\ee
where the last term comes from the perturbed shift vector.
Using \cref{eq:Carter-ids}, we obtain
\be 
	p_{ab} = \tfrac{1}{2\sqrt{2}} \lb[ - r_ar_b P_1 + \sqrt{2} r_{(a}m_{b)} P_2 + \sqrt{2} r_{(a}\bar{m}_{b)}P_3 + m_a m_b P_4 + m_{(a} \bar{m}_{b)} P_5  \rb] 
	\label{eq:p-hertz}
\ee
where
\begin{subequations}\label{eq:Ps-defn}\begin{align}
	P_1 & \defn 4 U \rho + \eth V \label{eq:P1-defn} \\
    P_2 & \defn \eth'U - (\thorn + \rho)V = \eth' U - 2(\eth W + \rho V) - 4\rho^2 \eth \psi \label{eq:P2-defn}\\
    P_3 & \defn \eth U \label{eq:P3-defn} \\
    P_4 & \defn \eth' V - D_t W = - (D_r - 4\rho) (W + 2\rho\eta) \label{eq:P4-defn} \\
    P_5 & \defn (3\thorn - \thorn' - 4\rho) U - \eth V = (D_r - 4\rho) U \label{eq:P5-defn} 
\end{align}\end{subequations}
Here $W$ is viewed as the properly GHP-weighted scalar defined by \cref{eq:W-used-Teu1}, (as opposed to the formula \cref{eq:W-used-Teu2}, which holds only in the Carter frame), so that $\eth W$ is well defined.
Note that the last form of these quantities are expressed in terms of derivatives intrinsic to \(\Sigma\) and can be written in terms of the initial data \((\psi,\eta)\vert_\Sigma\) using \cref{eq:U-V-W-alt}. Note also that $P_2$, $P_4$, and $P_5$ do not have well defined GHP-weights, since they are sums of terms with different GHP-weights.

As discussed at the end of \cref{sec:Teu-adj-hertz} (see \cref{eq:hertz-smooth}), we require that $\psi = \Delta \tilde\psi$, where $\tilde \psi$ is smooth at $B$. This implies conditions \cref{eq:U-V-W-smooth} on $U,V, W$. These conditions yield, in turn (with \(\delta\vartheta_{\pm} = \delta\vartheta_{\rsub{odd}} \pm \delta\vartheta_{\rsub{even}}\))
\begin{subequations}\begin{align}
	\delta \varepsilon & = \lb. \tfrac{1}{2} s^{ab} q_{ab} \rb\vert_B = 0  \label{eq:pert-area}\\
	\delta \vartheta_{\rsub{odd}} & = \lb. -r^a r^b p_{ab}  \rb\vert_B = \lb. -\tfrac{1}{2\sqrt{2}} (4\rho U + \eth V) \rb\vert_B = 0 \label{eq:pert-exp-odd}\\
	\delta \vartheta_{\rsub{even}} & = \lb. \tfrac{1}{2} \lb[ r^a D_a (s^{bc} q_{bc}) - 2 s^{ab} \ms D_a (q_{b c}r^c) \rb] \rb\vert_B = \lb. \tfrac{1}{2\sqrt{2}} \eth V  \rb\vert_B = 0 \label{eq:pert-exp-even}
\end{align}\end{subequations}
Thus, the conditions \cref{eq:conditions} are automatically satisfied for perturbations generated by a Hertz potential satisfying \cref{eq:hertz-smooth}. Every perturbation generated by a Hertz potential also satisfies
\be
	\delta M = \delta J = \delta P_i = 0
\ee
Near spatial infinity, we further require that the initial data for the Hertz potential satisfies $\psi= O(r^{\nfrac{1}{2} + \epsilon})$, \(\eta = O(1/r^{\nfrac{1}{2} + \epsilon})\)
and all \(n\)\textsuperscript{th} spatial derivatives of \(\psi\) and \(\eta\) fall off by an additional factor of \(1/r^n\). Using \cref{eq:U-V-W-alt} this implies that we have $U, V, W = O(1/r^{\nfrac{3}{2} + \epsilon})$. It then follows that the metric perturbation generated by $\psi$ lies in $\ms V_c^\infty$.

Without loss of generality\footnote{One way of seeing that this involves no loss of generality is to note that positivity of canonical energy will hold if and only if it holds separately for each $(\ell,m)$ in a spherical harmonic expansion. By rotational invariance, for a given $\ell$ the canonical energy cannot depend on $m$, so it suffices to consider only the case $m=0$.} for analyzing the positivity of canonical energy, we will assume in the following that the Hertz potential \(\psi\) --- and thus, the metric perturbation generated by \(\psi\) --- is axisymmetric. In addition to the \(t\)-\(\phi\) reflection isometry possessed by all stationary-axisymmetric black holes \cite{SW-tphi}, Schwarzschild spacetime is static and possesses separate $t$ and $\phi$ reflection isometries. 
Since the \(t=0\) Cauchy surface is invariant under the \(\phi\)-reflection isometry, we can decompose the initial data for the metric perturbations into a sum of parts that are reflection odd (\emph{axial}) and reflection even (\emph{polar}) under this isometry. Since the canonical energy is invariant under the action of the reflection isometry, there cannot be any ``cross-terms'' arising from the axial and polar contributions to $\ms E$, i.e., we have $\ms E =\ms E^{\rsub{(axial)}} + \ms E^{\rsub{(polar)}}$, where $\ms E^{\rsub{(axial)}}$ and $\ms E^{\rsub{(polar)}}$ are, respectively, the canonical energies of the axial and polar parts of the perturbation.

We now compute the kinetic and potential energies of axial and polar initial data in terms of $\psi$ and $\eta$. In fact, by the general proof of \cite{PW}, we know that the kinetic energy is always positive (on any stationary-axisymmetric background spacetime in any number of dimensions), but it will be useful to have the explicit kinetic energy expressions.
From \cref{eq:p-hertz,eq:q-hertz} the axial initial data are
\begin{subequations}\label{eq:pq-axial}\begin{align}
	p_{ab}^{\rm(axial)} & = \tfrac{1}{4\sqrt{2}} \lb[ \sqrt{2} r_{(a}(m_{b)} - \bar m_{b)}) (P_2 - P_3) + (m_a m_b - \bar m_a \bar m_b) P_4 \rb] \label{eq:p-axial} \\[1.5ex]
\begin{split}
	q_{ab}^{\rsub{(axial)}} & = - \tfrac{1}{2\sqrt{2}} V r_{(a} \lb( m_{b)} - \bar m_{b)} \rb) - \tfrac{1}{2}W \lb( m_a m_b - \bar m_a \bar m_b \rb) 
\end{split}\label{eq:q-axial} 
\end{align}\end{subequations}
Computing the kinetic energy from \cref{eq:complex-KE-defn} we get
\be\label{eq:K-axial}\begin{split}
	16\pi \ms K^{\rsub{(axial)}} & = \tfrac{1}{8}\int_\Sigma N \lb[ \abs{P_2 - P_3}^2 + \abs{P_4}^2 \rb] \\
	& = \tfrac{1}{8}\int_\Sigma N \lb[ \abs{(\thorn + \rho)V}^2 + \abs{D_t W - \eth' V }^2 \rb] 
\end{split}\ee
where we have used \cref{eq:Ps-defn} and \cref{eq:axisymm-id} for \(U\) in the second line.

Similarly, using \cref{eq:axisymm-id} for \(V\) and \(W\) we  can compute the potential energy from \cref{eq:complex-PE-defn}
\be\label{eq:U-axial1}\begin{split}
	16\pi \ms U^{\rsub{(axial)}} & = \tfrac{1}{8} \int_\Sigma N \abs{D_r W + \eth' V}^2 \\
	&\quad + \tfrac{1}{8} \int_\Sigma N \lb[\abs{\eth V}^2 - \abs{\eth' V}^2 - \rho D_r \abs{V}^2 \rb] \\
	&\quad + \int_\Sigma N \lb[ 2\beta \eth \abs{W}^2 - \tfrac{1}{2} \rho D_r \abs{W}^2 \rb]
\end{split}\ee
Converting to the NP derivatives, using \cref{eq:axisymm-id,eq:IBP} for both \(V\) and \(W\), we see that the last two lines of \cref{eq:U-axial1} vanish. Thus, we obtain
\be\label{eq:U-axial}
	16\pi \ms U^{\rsub{(axial)}} = \tfrac{1}{8} \int_\Sigma N \abs{D_r W + \eth' V}^2
\ee
From \cref{eq:K-axial,eq:U-axial} we see that the canonical energy \(\ms E^{\rsub{(axial)}} = \ms K^{\rsub{(axial)}} + \ms U^{\rsub{(axial)}}\) of the axial part of the complex metric perturbation is manifestly positive.\footnote{This positivity result for axial perturbations also could also have been shown directly from the expressions \cref{eq:complex-KE-defn,eq:complex-PE-defn}, without the need to introduce Hertz potentials.}

The polar initial data are
\begin{subequations}\label{eq:pq-polar}\begin{align}
\begin{split}
	p_{ab}^{\rsub{(polar)}} & = \tfrac{1}{4\sqrt{2}} \big[ - r_ar_b 2 P_1 + \sqrt{2} r_{(a}( m_{b)} + \bar m_{b)} ) (P_2 + P_3) \\
		&\qquad + (m_a m_b + \bar m_a \bar m_b) P_4 + m_{(a} \bar{m}_{b)} 2 P_5 \big]
\end{split} \label{eq:p-polar} \\[1.5ex]
	q_{ab}^{\rsub{(polar)}} & = -\tfrac{1}{2} r_a r_b U - \tfrac{1}{2\sqrt{2}} V r_{(a}( m_{b)} + \bar m_{b)} ) -\tfrac{1}{2} W (m_a m_b + \bar m_a \bar m_b) \label{eq:q-polar} 
\end{align}\end{subequations}

Using the identities \cref{eq:axisymm-id,eq:IBP} for \(U,V,W\) we obtain
\begin{subequations}\label{eq:KU-polar}\begin{align}
	16\pi \ms K^{\rsub{(polar)}} & = 16\pi \ms K^{\rsub{(axial)}} + \tfrac{1}{8} \int_\Sigma N \lb[ \abs{P_1}^2 + 2 (\bar P_2 P_3 + P_2 \bar P_3) + (\bar P_1 P_5 + P_1 \bar P_5) \rb] \label{eq:K-polar} \\[1.5ex]
\begin{split}\label{eq:U-polar}
	16\pi \ms U^{\rsub{(polar)}} & = 16\pi \ms U^{\rsub{(axial)}} - \tfrac{1}{8} \int_\Sigma N \bigg[ \abs{\eth V}^2 + D_r U \eth' \bar{V} + D_r \bar{U} \eth V  \\
	&\quad + 4 \abs{\eth U}^2 + 4 \rho D_r \abs{U}^2 - 4 \eth' \bar{U} (\eth W + \rho V) - 4 \eth U (\eth' \bar{W} + \rho \bar V) \bigg]
\end{split}
\end{align}\end{subequations}
It may not appear obvious from \cref{eq:KU-polar} that the canonical energy \(\ms E^{\rsub{(polar)}} = \ms K^{\rsub{(polar)}} + \ms U^{\rsub{(polar)}}\) of polar perturbations is positive. However, we now show that \(\ms E^{\rsub{(polar)}} = \ms E^{\rsub{(axial)}}\), i.e., the canonical energies of the polar and axial perturbations are in fact equal, and hence \(\ms E^{\rsub{(polar)}}\) is also positive.

Using \cref{eq:Ps-defn}, we can write the terms in the integrand of the last term in \cref{eq:K-polar} as
\begin{subequations}\begin{align}
	\abs{P_1}^2 + (\bar P_1 P_5 + P_1 \bar P_5) & = \abs{\eth V}^2 + 4\rho D_r \abs{U}^2 + \eth' \bar V D_r U + \eth V D_r \bar U - 16\rho^2 \abs{U}^2 \\
\begin{split}
	2 (\bar P_2 P_3 + P_2 \bar P_3) & = -4 \eth'\bar U (\eth W + \rho V) -4 \eth U (\eth'\bar W + \rho \bar V) + 4 \abs{\eth U}^2 \\
		&\quad - 8 \rho^2 \eth'\bar U \eth \psi - 8 \rho^2 \eth U \eth' \bar\psi
\end{split}
\end{align}\end{subequations}
where we have used \cref{eq:axisymm-id} for \(U\). Using these in \cref{eq:K-polar,eq:U-polar} we get
\be\begin{split}
	16\pi (\ms E^{\rsub{(polar)}} - \ms E^{\rsub{(axial)}}) & =  \int_\Sigma N [- 16\rho^2 \abs{U}^2 - 8 \rho^2 \eth'\bar U \eth \psi - 8 \rho^2 \eth U \eth' \bar\psi ]\\
	&= -8 \int_\Sigma N \rho^2 [ \eth(\bar U \eth \psi) + \eth'(U \eth' \bar\psi) ] \\
	&=  -8 \int_\Sigma N  (\delta + 2\beta)(\rho^2 \bar U \eth \psi + \rho^2 U \eth' \bar\psi ) 
\end{split}\ee
where the second line used \cref{eq:axisymm-id} for \(U\). Finally, the last line vanishes due to \cref{eq:IBP-sphere}, so we have \(\ms E^{\rsub{(polar)}} = \ms E^{\rsub{(axial)}}\). 
Thus, the full canonical energy of the metric perturbation generated by a Hertz potential is \(\ms E = \ms E^{\rsub{(polar)}} + \ms E^{\rsub{(axial)}} = 2 \ms E^{\rsub{(axial)}}\), so
\be\label{eq:energy-Hertz}
	\ms E = \tfrac{1}{64\pi}\int_\Sigma N \lb[ \abs{(\thorn + \rho)V}^2 + \abs{D_t W - \eth' V }^2 + \abs{D_r W + \eth' V}^2 \rb] 
\ee
which is manifestly positive.

So far we have shown that the canonical energy of the \emph{complex} perturbation \(\gamma_{ab}\) generated by a Hertz potential is positive. However, we are interested in the canonical energy of the \emph{real} perturbation \(\Re\gamma_{ab}\) generated by a Hertz potential. However, the positivity of $\ms E(\Re\gamma_{ab})$ will follow as a consequence of the following lemma:

\begin{lemma} \label{symorth} Let $\psi$, $\psi'$ be Hertz potentials and let $\gamma = \adj{\mc S}[\psi]$, \(\gamma' = \adj{\mc S}[\psi']\) be the corresponding complex metric perturbations generated by these Hertz potentials. Then
\be
\Omega(\gamma, \gamma') = 0
\ee
\begin{proof}
By direct substitution of \cref{eq:q-hertz,eq:p-hertz} into \cref{Omega2}, we obtain
\be
    \Omega(\gamma, \gamma') = \tfrac{1}{4\sqrt{2}}\int_\Sigma \eth (U V' - U'V) = \tfrac{1}{4\sqrt{2}}\int_\Sigma (\delta + 2\beta)(U V' - U'V) = 0
\ee
where \cref{eq:IBP-sphere} was used.
\end{proof}
\end{lemma} 

\begin{remark}
It follows as an immediate consequence of this lemma and \cref{eq:Omega-Pi} that for any \emph{complex} metric perturbation $\gamma$ generated by a Hertz potential, we have
\be
\mc T[\gamma] = 0
\ee
Thus, for any complex metric perturbation generated by a Hertz potential, the perturbed Weyl tensor component $\psi_0$ vanishes, i.e., the contributions of the real and imaginary parts of the metric perturbation to $\psi_0$ cancel. However, it can be verified that for non-stationary perturbations, we have $\psi_4 \neq 0$, and the complex metric perturbation $\gamma$ does \emph{not} give rise to a self-dual perturbed Weyl tensor \cite{Wald-sd}.\footnote{Complex electromagnetic perturbations generated by an electromagnetic Hertz potential do give rise to a self-dual Maxwell field tensor (see \cref{Fhertz} below).}
\end{remark}

The relevance of \cref{symorth} can be seen as follows. Taking $\gamma' = \Lie_t \gamma$ (so that $\gamma'$ is the complex metric perturbation generated by the Hertz potential $\Lie_t \psi$), we obtain
\begin{eqnarray}
0 &=& \Omega(\gamma, \gamma') = \Omega(\Re \gamma + i \Im \gamma, \Lie_t \Re \gamma + i \Lie_t \Im \gamma) \nonumber \\
&=& \ms E(\Re \gamma, \Re \gamma) - \ms E(\Im \gamma, \Im \gamma) + 2 i \ms E(\Re \gamma, \Im \gamma)
\end{eqnarray}
where the bi-linearity of $\Omega$ and the symmetry of $\ms E$ for real metric perturbations were used.
The real part of this equation yields
\be
\ms E(\Re \gamma) = \ms E(\Im \gamma)
\ee
On the other hand, since we defined $\ms E$ so that it is a Hermitian form on complex perturbations, it is easily seen that 
\be
\ms E (\gamma) = \ms E(\Re \gamma) + \ms E(\Im \gamma)
\ee
Thus, we have
\be
    \ms E(\Re \gamma) = \tfrac{1}{2} \ms E(\gamma) \geq 0
\ee 
Thus, the canonical energy of any real perturbation of Schwarzschild generated by a Hertz potential is positive.

\section{Relationship of the energy in the DHR analysis to the canonical energy of an associated perturbation}\label{sec:DHR-compare}

In this section we relate the energy quantity underlying the results of Dafermos, Holzegel and Rodnianski (DHR) \cite{DHR} to the Hertz potential and canonical energy. 

We start by briefly reviewing the variables used by \cite{DHR}, translating to our notation. The frame \(\mc N_{EF}\) used by \cite{DHR} corresponds (up to numerical constants) to the Carter frame \cref{eq:Carter-frame}.  Let \(\hat\gamma_{ab}\) be a perturbation of Schwarzschild as considered by \cite{DHR} and let \(\hat\psi_4\) denote the perturbation in the Weyl scalar \(\Psi_4 \defn - C_{abcd}n^a \bar m^b n^c \bar m^d \). The real, symmetric, traceless tensor Regge-Wheeler variable, $\underline\Psi_{ab}$, of \cite{DHR} --- defined in Eq.~195 of their paper --- is given in terms of the Teukolsky variable \(\hat\psi_4\) by\footnote{Note that $\hat\psi_4 m_a m_b$ corresponds (up to numerical factors) to the quantity \(\overset{\rsub{(1)}}{\underline\alpha}_{ab}\) in \cite{DHR}.}
\be\label{eq:DHR-var-tensor}
	\underline\Psi_{ab} \defn -\frac{r^3}{4\sqrt{\Delta}}D \lb[ \frac{r^3}{\sqrt{\Delta}}D \lb\{ \frac{\Delta}{r} \lb( m_a m_b \hat\psi_4 + \bar m_a \bar m_b \bar{\hat\psi}_4 \rb) \rb\} \rb]
\ee
This can be re-written in terms of a complex scalar DHR variable $\Psi_{\rsub{DHR}}$ as
\be\label{eq:DHR-tensor-scalar}
	\underline\Psi_{ab} = -\frac{r}{4} \lb( m_a m_b \Psi_{\rsub{DHR}} + \bar m_a \bar m_b \bar\Psi_{\rsub{DHR}} \rb)
\ee
where
\be\label{eq:DHR-var-Hertz}
	\Psi_{\rsub{DHR}} = (\thorn + \rho)(\thorn + 3\rho)\psi \wt (-2,2)
\ee
and we have written\footnote{For Petrov type-D spacetimes, the rescaling factor from \(\hat\psi_4\) to \(\psi\) is given by the coefficient of the \emph{unique} Killing spinor in the principal null frame \cite{AA,Aks}. Up to constant factors this is equivalent to rescaling by \((-\Psi_2)^{-4/3}\), or by \((-\rho)^{-4}\) in the Kinnersley frame. For the Schwarzschild case this corresponds to a rescaling by \(r^4\) as in \cref{eq:Weyl-to-Hertz}.}
\be\label{eq:Weyl-to-Hertz}
\psi = r^4 \hat\psi_4 \wt (-4,0)
\ee

The DHR variables \(\underline\Psi_{ab}\) and \(\Psi_{\rsub{DHR}}\) satisfy, respectively, a tensor and spin-weighted Regge-Wheeler equation \cite{DHR} (with \(\ms D^2 = s^{ab} \ms D_a \ms D_b\))
\begin{subequations}\label{eq:DHR-RW-eqn}\begin{align}
	2\frac{\sqrt{\Delta}}{r} D' \lb(\frac{\sqrt{\Delta}}{r} D \underline\Psi_{ab}\rb) + \frac{\Delta}{r^2} \ms D^2 \underline\Psi_{ab} +  \frac{\Delta}{r^2} \mc V_{\rsub{DHR}} \underline\Psi_{ab} & = 0 \\[1.5ex]
	\bigg[ 2(\thorn' - \bar\rho')(\thorn - \rho) - 2\eth\eth' +\mc V_{\rsub{DHR}} \bigg] \Psi_{\rsub{DHR}} & = 0
\end{align}\end{subequations}
where the potential is given by
\be\label{eq:DHR-potential}
	\mc V_{\rsub{DHR}} = \frac{4}{r^2} - \frac{6M}{r^3}
\ee
The form of \cref{eq:DHR-RW-eqn} immediately implies that their solutions have a positive conserved energy given by\footnote{Recall that we are using a negative definite metric on $\Sigma$, so the first form of the energy in \cref{eq:DHR-energy} is also manifestly positive.} 
\be\label{eq:DHR-energy}\begin{split}
	\ms E_{\rsub{DHR}} & = \int_\Sigma N r^{-2} \bigg[ \tfrac{1}{2}\lb(D_t{\underline\Psi}_{ab} \rb)^2 + \tfrac{1}{2}  \lb( D_r\underline\Psi_{ab} \rb)^2 - \lb( \ms D_c \underline\Psi_{ab} \rb)^2 + \mc V_{\rsub{DHR}} \lb( \underline\Psi_{ab} \rb)^2 \bigg] \\
	& = \tfrac{1}{8}\int_\Sigma N~\bigg[ \tfrac{1}{2} \abs{ D_t\Psi_{\rsub{DHR}}}^2 + \tfrac{1}{2} \abs{ (D_r - 2\rho)\Psi_{\rsub{DHR}} }^2 + \abs{\eth\Psi_{\rsub{DHR}}}^2 + \abs{\eth'\Psi_{\rsub{DHR}}}^2 + \mc V_{\rsub{DHR}} \abs{\Psi_{\rsub{DHR}}}^2 \bigg]
\end{split}\ee
The flux of the corresponding energy on null hypersurfaces is evaluated by \cite{DHR} and used to obtain boundedness and decay results for the variable \(\underline{\Psi}_{ab}\). This is then used to show that the metric perturbation \(\hat\gamma_{ab}\) decays suitably to a perturbation towards a Kerr solution.

We now show that the energy \cref{eq:DHR-energy} can be obtained in a natural way, using Hertz potentials and canonical energy. Again, we start with a metric perturbation  \(\hat\gamma_{ab}\) as considered by \cite{DHR} and we obtain the corresponding Teukolsky variable \(\hat\psi_4\). But we now use $\psi =  r^4 \hat\psi_4$ as a \emph{Hertz potential} to generate a \emph{new, complex} metric perturbation $\gamma = \adj{\mc S}[\psi]$. We claim that 
\be
\ms E_{\rsub{DHR}} [\hat{\gamma}] = 4\pi \ms E [\gamma]
\ee
where $\ms E [\gamma]$ is the canonical energy of $\gamma$ (\cref{eq:energy-Hertz}). 

To show this, we note first that using \cref{eq:Teu-initial,eq:Teu-W,eq:Carter-ids1}, we have the following relations on \(\Sigma\) in the Carter frame
\begin{subequations}\label{eq:DHR-rels}\begin{align}
	\Psi_{\rsub{DHR}} & = D_r\eta + (\eth'\eth + 3\Psi_2)\psi = W + 2\rho \eta \\
	\eth'V & = 2\thorn \eth'\eth \psi \\
	(\thorn + \rho)V & = 2\eth \Psi_{\rsub{DHR}} \label{eq:V-DHR}
\end{align}\end{subequations}
Using Eqs.~\ref{eq:Carter-ids1}, \ref{eq:DN-epsilonrho}, \ref{eq:U-V-W-alt}--\ref{eq:W-used-Teu2} and \ref{eq:DHR-rels} we obtain
\be\label{eq:DHR-UVW}
    D_r W + \eth' V = D_t\Psi_{\rsub{DHR}} \eqsp D_t W - \eth' V = (D_r - 4\rho)\Psi_{\rsub{DHR}}
\ee
Thus, we can write the canonical energy \cref{eq:energy-Hertz} in terms of the DHR variable \(\Psi_{\rsub{DHR}}\) as
\be\label{eq:energy-compare1}
	16\pi \ms E = \tfrac{1}{2}\int_\Sigma N \lb[ \tfrac{1}{2} \abs{D_t\Psi_{\rsub{DHR}}}^2 + \tfrac{1}{2} \abs{(D_r - 4\rho)\Psi_{\rsub{DHR}}}^2 + 2 \abs{\eth\Psi_{\rsub{DHR}}}^2 \rb] 
\ee
Now, using \cref{eq:axisymm-id,eq:IBP} for \(\Psi_{\rsub{DHR}}\), we get
\begin{subequations}\begin{align}
	\begin{split}
	\int_\Sigma N \abs{\eth'\Psi_{\rsub{DHR}}}^2 
		& = \int_\Sigma N \lb[ \abs{\eth \Psi_{\rsub{DHR}}}^2 + 4(-\rho^2 + \Psi_2)\abs{\Psi_{\rsub{DHR}}}^2 \rb]
	\end{split} \\
	\begin{split}
	\int_\Sigma N \abs{(D_r - 4\rho)\Psi_{\rsub{DHR}}}^2 
	& = \int_\Sigma N \lb [\abs{(D_r - 2\rho)\Psi_{\rsub{DHR}}}^2 + (8\rho^2 + 4\Psi_2) \abs{\Psi_{\rsub{DHR}}}^2 \rb]
\end{split}
\end{align}\end{subequations}
where from \cref{eq:DHR-rels}, \(\Psi_{\rsub{DHR}} = o(1/r^{\nfrac{3}{2}})\) near spatial infinity. Using the above in \cref{eq:energy-compare1}, noting that \(\mc V_{\rsub{DHR}} = 8\rho^2 - 2\Psi_2\) from \cref{eq:DHR-potential,eq:Carter-spin,eq:Carter-curv} and comparing to \cref{eq:DHR-energy}, we have
\be\label{eq:energy-compare2}\begin{split}
	\ms E [\gamma] & = \tfrac{1}{32\pi}\int_\Sigma N \lb[ \tfrac{1}{2} \abs{D_t\Psi_{\rsub{DHR}}}^2 + \tfrac{1}{2} \abs{(D_r - 2\rho)\Psi_{\rsub{DHR}}}^2 + \abs{\eth\Psi_{\rsub{DHR}}}^2 + \abs{\eth'\Psi_{\rsub{DHR}}}^2 + \mc V_{\rsub{DHR}} \abs{\Psi_{\rsub{DHR}}}^2 \rb] \\
	& = \tfrac{1}{4\pi} \ms E_{\rsub{DHR}} [\hat{\gamma}]
\end{split}\ee

It is worth noting that since the Hertz potential construction generalizes to Kerr, we can construct an analogue of $\ms E_{\rsub{DHR}}$ for an axisymmetric metric perturbation, $\hat{\gamma}_{ab}$, of Kerr. Namely, we calculate the Teukolsky variable of $\hat{\gamma}_{ab}$ and then use it as a Hertz potential to generate a new, complex metric perturbation, $\gamma_{ab}$. We then compute the canonical energy, $\ms E [\gamma]$, of $\gamma_{ab}$. However, we have not shown that $\ms E [\gamma]$ must be positive for Kerr.

\section{Relation of the DHR variable to the Regge-Wheeler and twist potential variables of the associated perturbation}
\label{sec:RW-twist-compare}

In this section, we obtain relations between the DHR variable $\Psi_{\rsub{DHR}}$ associated with a metric perturbation $\hat{\gamma}_{ab}$, and the usual Regge-Wheeler variable for the axial part of the complex metric perturbation $\gamma_{ab}$ generated by a Hertz potential $\psi = r^4 \hat{\psi}_4$. We also relate $\Psi_{\rsub{DHR}}$ to the twist potential of $\gamma_{ab}$.

The original definition of the Regge-Wheeler variable given by \cite{RW} used a spherical harmonic expansion and a particular choice of gauge for the metric perturbation. This is not convenient for our purposes. Instead, we will use the gauge-invariant definition given by Moncrief \cite{Moncrief}, which does not require a spherical harmonic expansion. 
For any axial perturbation, the initial data \((p_{ab}^{\rsub{(axial)}}, q_{ab}^{\rsub{(axial)}})\) on the Cauchy surface \(\Sigma\) can be written in the form 
\be\label{eq:pq-axial-Mon}\begin{split}
	p_{ab}^{\rsub{(axial)}} & = 2 r_{(a} {\varepsilon_{b)}}^c\ms  D_c p_1 - 2{\varepsilon_{(a}}^c\ms D_{b)}\ms D_c p_2 \\
	q_{ab}^{\rsub{(axial)}} & = 2 r_{(a} {\varepsilon_{b)}}^c\ms  D_c q_1 - 2{\varepsilon_{(a}}^c\ms D_{b)}\ms D_c q_2
\end{split}\ee
where \(\varepsilon_{ab}\) is the volume form on the \(2\)-spheres, and $p_1, p_2, q_1, q_2$ are functions on \(\Sigma\). The Hamiltonian constraint \cref{eq:lin-Ham-constraint} is automatically satisfied for perturbations of this form, whereas the 
momentum constraint \cref{eq:lin-mom-constraint} determines the variable \(p_2\) in terms of \(p_1\) and its radial derivatives \cite{Moncrief}. 
Following \cite{Moncrief} we define the \emph{Regge-Wheeler variable} \(Q_{\rsub{RW}}\) and its conjugate momentum \(P_{\rsub{RW}}\), which are unconstrained gauge-invariant variables, as
\be\label{eq:RW-pq}\begin{split}
	Q_{\rsub{RW}} & \defn \sqrt{\Delta}\lb( \frac{q_1}{r^2} + \tfrac{\sqrt{\Delta}}{r}\partial_r \frac{q_2}{r^2} \rb) \\
	P_{\rsub{RW}} & \defn \tfrac{2r}{\Delta} \ms D^2 p_1 =  -\tfrac{4r}{\Delta} (\delta + 2\beta )\delta p_1
\end{split}\ee
Then \(Q_{\rsub{RW}}\) satisfies the scalar Regge-Wheeler equation \cite{RW,Moncrief}
\be\label{eq:RW-eqn}
	2\frac{\sqrt{\Delta}}{r} D' \lb(\frac{\sqrt{\Delta}}{r} D Q_{\rsub{RW}} \rb) + \frac{\Delta}{r^2} \ms D^2 Q_{\rsub{RW}} - \frac{\Delta}{r^2} \frac{6M}{r^3} Q_{\rsub{RW}} = 0
\ee

Now consider the complex metric perturbation \(\gamma_{ab}\) generated by the Hertz potential $\psi = r^4 \hat{\psi}_4$, as considered in the previous section. By comparing the form \cref{eq:pq-axial} of the axial part of this perturbation with \cref{eq:pq-axial-Mon} (where now $p_1, p_2, q_1, q_2$ are complex functions), and using \(\varepsilon_{ab} = 2 i m_{[a} \bar m_{b]}\), we find
\be\label{eq:RW-to-Hertz}\begin{split}
	\delta p_1 = \tfrac{i}{8} (\thorn + \rho)V  & \eqsp (\delta - 2\beta)\delta p_2 = \tfrac{i}{8\sqrt{2}} (D_t W - \eth'V) \\
	\delta q_1 = \tfrac{i}{4\sqrt{2}} V & \eqsp (\delta - 2\beta) \delta  q_2 = \tfrac{i}{4}W
\end{split}\ee
Using \cref{eq:RW-pq,eq:RW-to-Hertz,eq:V-DHR,eq:DHR-UVW}, the (complex) Regge-Wheeler variables for \(\gamma_{ab}\) satisfy
\be\label{eq:RW-to-DHR}\begin{split}
	(\delta - 2\beta)\delta Q_{\rsub{RW}} & = \frac{i\sqrt{\Delta}}{4\sqrt{2}r^2} \lb[ D_r W +\eth' V  \rb] = \frac{i\sqrt{\Delta}}{4\sqrt{2}r^2} D_t \Psi_{\rsub{DHR}} \\
	P_{\rsub{RW}} & = -i \frac{r}{2\Delta} \eth (\thorn + \rho)V = -i \frac{r}{\Delta} \eth^2 \Psi_{\rsub{DHR}}
\end{split}\ee
which gives the desired relation.

Another relation can be obtained by noting that the DHR variables can be written as (see Remark~7.1 \cite{DHR})
\begin{subequations}\begin{align}
	\tfrac{1}{r^2} \underline{\Psi}_{ab} & = \ms D_a \ms D_b f - \tfrac{1}{2}s_{ab} \ms D^2 f - \varepsilon_{(a}{}^c \ms D_{b)}\ms D_c g \label{eq:DHR-tensor-fg} \\
	\Psi_{\rsub{DHR}} & = -4 r(\delta - 2\beta) \delta (f + i g) \label{eq:DHR-scalar-fg}
\end{align}\end{subequations}
where \(f,g\) are uniquely determined by \(\Psi_{\rsub{DHR}}\) up to \(\ell = 0,1\) spherical harmonic modes and both \(f,g\) satisfy \cref{eq:RW-eqn}. Using \cref{eq:RW-to-DHR} we obtain
\be\label{eq:RW-fg}
	Q_{\rsub{RW}} = - i \Lie_t (f + i g)
\ee

Using the above relations, one can check that the Hamiltonian obtained by Moncrief \cite{Moncrief} for the Regge-Wheeler variable \(Q_{\rsub{RW}}\) (after a spherical harmonic decomposition) is equivalent to both the canonical energy \(\ms E [\gamma]\) and the DHR energy \(\ms E_{\rsub{DHR}} [\hat{\gamma}]\).\\

Next, we consider the twist potential variable which can be defined in spacetime as described in \cite{Geroch}. For our purposes it will be more convenient to use the following \((3+1)\)-formulation by Moncrief \cite{Mon-twist}. The linearized constraint \cref{eq:lin-mom-constraint} implies that the axial momentum perturbation \(p^{\rsub{(axial)}}_{ab}\) satisfies
\be
	D^a (p^{\rsub{(axial)}}_{ab} \phi^b) = 0
\ee
Following \cite{Mon-twist}, this can be solved by introducing a (complex) \emph{perturbed twist potential} \(\omega\)
\be\label{eq:twist-potential}
	p^{\rsub{(axial)}}_{ab} \phi^b = \tfrac{1}{2}\Phi^{-1} \varepsilon_a{}^{bc}\phi_c D_b \omega = - \tfrac{1}{r \sin\theta} \lb[ \tfrac{1}{\sqrt{2}} \delta \omega ~r_a + \tfrac{1}{4}D_r \omega ~(m_a + \bar m_a) \rb]
\ee
where \(\Phi \defn - \phi_a \phi^a = r^2\sin^2\theta\).

Using \cref{eq:p-axial,eq:Ps-defn,eq:V-DHR,eq:DHR-UVW} we find that for the complex metric perturbation generated by the Hertz potential $\psi = r^4 \hat{\psi}_4$, we have
\be\label{eq:p-axial-DHR}
	p^{\rsub{(axial)}}_{ab} \phi^b = - \tfrac{i}{8} r \sin\theta \lb[ \tfrac{1}{\sqrt{2}} 4 \eth \Psi_{\rsub{DHR}} ~ r_a + (D_r - 4\rho)\Psi_{\rsub{DHR}} (m_a + \bar m_a) \rb]
\ee
Converting the GHP derivatives to NP derivatives in \cref{eq:p-axial-DHR} and comparing with \cref{eq:twist-potential} yields
\be\label{eq:DHR-twist}
	\omega = \tfrac{i}{2} r^2\sin^2\theta~ \Psi_{\rsub{DHR}} 
\ee
Thus, $\Psi_{\rsub{DHR}}$ is simply related to the twist potential of the axial part of the complex metric perturbation generated from the Hertz potential $\psi = r^4 \hat{\psi}_4$.

\section*{Acknowledgements}
We thank Lars Andersson for suggesting to us that it should be useful to express the canonical energy in terms of a Hertz potential.
K.P. is supported in part by the NSF grants PHY-1404105 and PHY-1707800 to Cornell University. R.M.W. is supported in part by NSF grants PHY~15-05124 and PHY18-04216 to the University of Chicago. Some calculations used the computer algebra
system \textsc{Mathematica}~\cite{Mathematica}, in combination with the \textsc{xAct/xTensor} suite~\cite{xact,xact-spinors}.

\appendix

\section{Electromagnetic perturbations on Schwarzschild background}\label{sec:EM}

In this appendix, we will show the relationship of the energy obtained by Pasqualotto \cite{Pasq} for electromagnetic perturbations of Schwarzschild to the canonical energy of electromagnetic perturbations generated by a corresponding Hertz potential, in close analogy with the gravitational case treated in \cref{sec:DHR-compare}. We note that positivity of the canonical energy of electromagnetic perturbations of any static black hole (\emph{not} necessarily satisfying Einstein equation and in any number of dimensions) is easily shown \cite{PW-em}. Positivity of the canonical energy of electromagnetic perturbations of an arbitrary stationary-axisymmetric black hole solution of the vacuum Einstein equation was proven in \cite{PW-em}.

The operators for electromagnetic perturbations corresponding to the operators $\mc E, \mc T, \mc O, \mc S$ of \cref{sec:hertz} are \cite{Teu-more, Chrz, Wald-potential, CK-em}
\be\begin{split}
	\mc E_{\rsub{EM}} [\hat{\ms A}_a] & = 2 \nabla^b \nabla_{[b} \hat{\ms A}_{a]}  \\
	\mc T_{\rsub{EM}} [\hat{\ms A}_a ] & \defn \hat\varphi_0 \wt (2,0)\\
	\mc O_{\rsub{EM}}[ \hat\varphi_0] &\defn \lb[ \lb( \thorn - 2\rho - \bar\rho \rb) \lb( \thorn' - \rho' \rb) - \lb( \eth + 2\tau - \bar\tau' \rb)\lb( \eth' - \tau' \rb) \rb] \hat\varphi_0 \wt (2,0) \\
	2\mc S_{\rsub{EM}} [J_a] & \defn \lb( \eth - 2\tau - \bar\tau' \rb)(l^aJ_a) - \lb( \thorn - 2\rho - \bar\rho \rb)(m^aJ_a) \wt (2,0)
\end{split}\ee
These operators again satisfy the identity \cref{eq:SEOT}, and their adjoints therefore also satisfy \cref{eq:TOES}. 
It is easily seen that $\mc E_{\rsub{EM}}$ is self-adjoint. The adjoints of \(\mc O_{\rsub{EM}}\) and \(\mc S_{\rsub{EM}}\) are
\be\begin{split}
	\adj{\mc O}_{\rsub{EM}} [\varphi] & \defn \lb[ \lb( \thorn' - \bar\rho' \rb)\lb( \thorn + \rho \rb) - \lb( \eth' - \bar\tau \rb)\lb( \eth + \tau \rb) \rb] \varphi \wt (-2,0) \\
	2\adj{\mc S}_{\rsub{EM}} [\varphi] & \defn  \lb[ - l^a \lb( \eth + \tau \rb) + m^a \lb( \thorn + \rho \rb) \rb]\varphi \wt (0,0)
\end{split}\ee
where \(\varphi \wt (-2,0)\). The equation \(\adj{\mc O}_{\rsub{EM}} [\varphi] = 0\) is just the spin-\((-1)\) Teukolsky equation. 
We may use solutions to  \(\adj{\mc O}_{\rsub{EM}} [\varphi] = 0\) as Hertz potentials to generate the \emph{complex} vector potential \(\ms A^a = 2\adj{\mc S}_{\rsub{EM}}[\varphi]\) solutions to Maxwell equation. Our choice of $\adj{\mc S}_{\rsub{EM}}$ puts the vector potential in the ``ingoing radiation'' gauge \(\ms A_a l^a = 0\). 

We now consider a vector potential $\ms A_a$ generated by the Hertz potential $\varphi$ on a Schwarzschild spacetime given by
\begin{subequations}\label{eq:A-Hertz-Sch}\begin{align}
	\ms A_a & = -l_a \eth \varphi + m_a \chi \quad\text{with} \label{eq:A-XY}\\[1.5ex]
	\begin{split}
		\chi &\defn (\thorn + \rho)\varphi \wt (-1,1)
\end{split} \label{eq:XY-defn}
\end{align}\end{subequations}
The spin-\((-1)\) Teukolsky equation \(\adj{\mc O}_{\rsub{EM}} [\varphi] = 0\) becomes
\be\label{eq:Teu-EM}
	\thorn'\chi = \bar\rho' \chi + \eth'\eth \varphi
\ee
Using \cref{eq:GHP-comm,eq:Dspin}, we obtain
\be\label{eq:XY-rel}
	\thorn \eth \varphi = \eth \chi
\ee

Using \cref{eq:Teu-EM,eq:XY-rel}, the complex Maxwell field strength \(\ms F_{ab} = 2 \nabla_{[a} \ms A_{b]}\) is computed to be
\be \label{Fhertz}
	\ms F_{ab} = 2 ( l_{[a} n_{b]} + m_{[a} \bar{m}_{b]} ) \eth \chi - 2 l_{[a} \bar{m}_{b]} \eth^2 \varphi + 2 n_{[a} m_{b]} (\thorn - \rho) \chi 
\ee
Note that $\ms F_{ab}$ is self-dual, i.e., \((* \ms F)_{ab} = i \ms F_{ab}\).

On the Cauchy surface \(\Sigma\), the initial data is given by the spatial vector potential \(A_a\) and the \emph{electric field} \(E_a\)
\be\label{eq:A-E-spatial}\begin{split}
	A_a & = \tfrac{1}{\sqrt{2}}r_a \eth \varphi + m_a \chi \\
	E_a & = r_a \eth \chi - \tfrac{1}{\sqrt{2}} \bar m_a \eth^2 \varphi + \tfrac{1}{\sqrt{2}} m_a (\thorn - \rho) \chi
\end{split}\ee
The \emph{magnetic field} is
\be\label{eq:F-spatial}\begin{split}
	F_{ab} = 2 D_{[a} A_{b]} = \sqrt{2} r_{[a} \bar{m}_{b]} \eth^2 \varphi + \sqrt{2} r_{[a} m_{b]} (\thorn - \rho) \chi + 2  m_{[a} \bar{m}_{b]} \eth \chi
\end{split}\ee
The canonical energy is\footnote{As discussed in \cite{PW-em}, the canonical energy differs from the ordinary electromagnetic energy by a boundary term at $B$. However, this boundary term vanishes for a static black hole.}
\be
\ms E_{\rsub{EM}} [\ms A] = \tfrac{1}{4\pi} \int_\Sigma N~ \lb[- \bar E_a E^a + \tfrac{1}{2} \bar F_{ab} F^{ab} \rb] 
\ee
Again, we remind the reader that the negative sign of the \(\bar E_a E^a\) term is due to the negative-definite spatial metric.
However, as a direct consequence of the self-duality of \(\ms F_{ab}\), we have
\be\begin{split}
	-\int_\Sigma N \bar E_a E^a = \tfrac{1}{2} \int_\Sigma N \bar F_{ab} F^{ab} 
\end{split}\ee
Using \cref{eq:A-E-spatial,eq:F-spatial} the canonical energy \(\ms E_{\rsub{EM}}\) can be put in the form
\be\label{eq:can-energy-EM}\begin{split}
	\ms E_{\rsub{EM}} [\ms A]& = \tfrac{1}{4\pi} \int_\Sigma N \lb[2 \abs{\eth \chi}^2 + \tfrac{1}{2} \abs{\eth^2 \varphi + (\thorn - \rho) \chi}^2 + \tfrac{1}{2} \abs{\eth^2 \varphi - (\thorn - \rho) \chi}^2 \rb] \\
		& = \tfrac{1}{4\pi} \int_\Sigma N \lb[ \tfrac{1}{2} \abs{D_t \chi }^2 + \tfrac{1}{2} \abs{(D_r - 2\rho)\chi }^2 + 2\abs{\eth \chi}^2 \rb]
\end{split}\ee
where to get the last line we have used \cref{eq:Teu-EM} and \cref{eq:axisymm-id} for \(\eth \varphi\).

We now relate the energy obtained by \cite{Pasq} to Hertz potentials and canonical energy. Let \(\hat{\ms A}_a\) be an electromagnetic perturbation and \(\hat\varphi_2 \defn \hat{\ms F}_{ab}\bar m^a n^b \wt (-2,0)  \) the corresponding perturbed Maxwell field strength component. Pasqualotto \cite{Pasq} considers the quantity
\be\label{eq:Pasq-var-vector}
	\underline\Phi_a \defn - \frac{\sqrt{2}r^3}{\sqrt{\Delta}}D \lb[ \sqrt{2\Delta} \lb( m_a \hat\varphi_2 + \bar m_a \bar{\hat\varphi}_2 \rb) \rb] 
\ee
On a Schwarzschild background, \(\varphi = r^2 \hat\varphi_2 \wt (-2,0) \) solves the spin-\((-1)\) Teukolsky equation \(\adj{\mc O}_{\rsub{EM}} [\varphi] = 0\), and can be used as a Hertz potential to generate electromagnetic perturbations. In terms of \(\varphi\) we can write
\be\label{eq:Pasq-vector-scalar}
	\underline{\Phi}_a = 2 r (m_a \Phi_{\rsub{P}} + \bar m_a \bar\Phi_{\rsub{P}})
\ee
with the complex scalar
\be\label{eq:Pasq-var-Hertz}
	\Phi_{\rsub{P}} = (\thorn + \rho)\varphi \wt (-1,1)
\ee
The Maxwell equation for \(\hat{\ms A}_a\) gives the relation \(\eth'\hat\varphi_1 = (\thorn-\rho)\hat\varphi_2\) \cite{GHP} and so we also have
\begin{subequations}\label{eq:phi1-Pasq}\begin{align}
	-\tfrac{1}{2r^3} \underline{\Phi}_a & = \ms D_a (\Re \hat\varphi_1) + \varepsilon_a{}^b \ms D_b (\Im \hat\varphi_1) \label{eq:phi1-Pasq-vector} \\
	\Phi_{\rsub{P}} & = r^2 \eth' \hat\varphi_1 \label{eq:phi1-Pasq-scalar}
\end{align}\end{subequations}

The variables \(\underline{\Phi}_a\) and \(\Phi_{\rsub{P}}\) satisfy, respectively, the vector and spin-weighted Fackerell-Ipser equation \cite{Pasq}
\be\label{eq:Pasq-FI-eqn}\begin{split}
	2 \frac{\sqrt{\Delta}}{r} D' \lb( \frac{\sqrt{\Delta}}{r} D \underline\Phi_a \rb) + \frac{\Delta}{r^2} \ms D^2\underline\Phi_a + \frac{\Delta}{r^2} \mc V_{\rsub{P}} \underline\Phi_a & = 0 \\
	\lb[ 2(\thorn' - \bar\rho')(\thorn - \rho) - 2\eth\eth' +\mc V_{\rsub{P}} \rb] \Phi_{\rsub{P}} & = 0
\end{split}\ee
with the potential
\be\label{eq:Pasq-potential}
	\mc V_{\rsub{P}} = \frac{1}{r^2}
\ee

The conserved energy for solutions of \cref{eq:Pasq-FI-eqn} is given by
\be\label{eq:Pasq-energy}\begin{split}
	\ms E_{\rsub{P}} [\hat{\ms A}] & = \int_\Sigma N r^{-2} \lb[- \tfrac{1}{2}\lb(D_t{\underline\Phi}_{a} \rb)^2 - \tfrac{1}{2} \lb( D_r\underline\Phi_{a} \rb)^2 + \lb( \ms D_b \underline\Phi_{a} \rb)^2 - \mc V_{\rsub{P}} \lb( \underline\Phi_{a} \rb)^2 \rb] \\
	& = \int_\Sigma N \lb[ \tfrac{1}{2} \abs{D_t \Phi_{\rsub{P}}}^2 + \tfrac{1}{2} \abs{ (D_r - 2\rho)\Phi_{\rsub{P}} }^2 + \abs{\eth\Phi_{\rsub{P}}}^2 + \abs{\eth'\Phi_{\rsub{P}}}^2 + \mc V_{\rsub{P}} \abs{\Phi_{\rsub{P}}}^2  \rb]
\end{split}\ee
This energy is in fact equivalent to the canonical energy \cref{eq:can-energy-EM}. To see this first note from \cref{eq:Pasq-var-Hertz,eq:XY-defn} we have, \(\chi = \Phi_{\rsub{P}}\). Thus, \cref{eq:can-energy-EM} is
\be\label{eq:energy-EM-compare1}
	\ms E_{\rsub{EM}} [\ms A] = \tfrac{1}{4\pi} \int_\Sigma N \lb[ \tfrac{1}{2} \abs{D_t \Phi_{\rsub{P}} }^2 + \tfrac{1}{2} \abs{(D_r - 2\rho)\Phi_{\rsub{P}} }^2 + 2\abs{\eth \Phi_{\rsub{P}}}^2 \rb]
\ee
Then, using \cref{eq:axisymm-id} for \(\Phi_{\rsub{P}}\), and \cref{eq:IBP-sphere,eq:Carter-spin,eq:Carter-curv}, we have
\be\begin{split}
	\int_\Sigma N \abs{\eth' \Phi_{\rsub{P}}}^2 
		& = \int_\Sigma N \lb[ \abs{\eth \Phi_{\rsub{P}}}^2 - \tfrac{1}{r^2} \abs{\Phi_{\rsub{P}}}^2  \rb]
\end{split}\ee
Using this in \cref{eq:energy-EM-compare1}, noting the form of the potential \cref{eq:Pasq-potential}, and comparing to \cref{eq:Pasq-energy}, we get
\be\label{eq:energy-EM-compare2}\begin{split}
	\ms E_{\rsub{EM}} [\ms A] & = \tfrac{1}{4\pi} \int_\Sigma N \lb[ \tfrac{1}{2} \abs{D_t \Phi_{\rsub{P}}}^2 + \tfrac{1}{2} \abs{ (D_r - 2\rho)\Phi_{\rsub{P}} }^2 + \abs{\eth\Phi_{\rsub{P}}}^2 + \abs{\eth'\Phi_{\rsub{P}}}^2 + \mc V_{\rsub{P}} \abs{\Phi_{\rsub{P}}}^2  \rb] \\
	& = \tfrac{1}{4\pi} \ms E_{\rsub{P}} [\hat{\ms A}]
\end{split}\ee
as we desired to show.

It is also worth clarifying the relationship between the form of the canonical energy \cref{eq:can-energy-EM} and the form obtained in \cite{PW-em,G-em}. We first note that since the field strength \(\ms F_{ab}\) is self-dual its charge vanishes. Then, consider the complex scalar variable
\be\label{eq:Ernst-potential}
	\Psi \defn \phi^a A_a = i \sqrt{2}r \sin\theta ~\chi = i \sqrt{2}r \sin\theta ~\Phi_{\rsub{P}}
\ee
which can be viewed as a ``magnetic'' potential for \(\ms F_{ab}\) (see Remark~5.1 \cite{PW-em}) or as an \emph{Ernst potential} \cite{Ernst1,Ernst2}. In terms of \(\Psi\) we can write the canonical energy \cref{eq:can-energy-EM} as
\be\label{eq:can-energy-EM3}
	\ms E_{\rsub{EM}} = \tfrac{1}{8\pi} \int_\Sigma N^{-1} \Phi^{-1} \abs{\dot\Psi }^2 - N \Phi^{-1} h^{ab} D_a\Psi D_b\bar\Psi
\ee
On \(\Sigma\), define the \(1\)-form \(Z_a\) by
\be
	Z_a \defn - \tfrac{1}{2} \Phi^{-1} D_a \Phi \equiv - \tfrac{1}{r}~ dr - \cot\theta~ d\theta 
\ee
where \(\Phi \defn - \phi_a \phi^a = r^2\sin^2\theta\). \(Z_a\) is related to the anti-self-dual Ashtekar-Sen connection of the background Schwarzschild spacetime, see Remark~5.2 of \cite{PW-em}. Integrating-by-parts in \cref{eq:can-energy-EM3} we obtain
\be
	\ms E_{\rsub{EM}} = \tfrac{1}{8\pi} \int_\Sigma N^{-1} \Phi^{-1} \abs{\dot\Psi }^2 - N \Phi^{-1} h^{ab}\lb[ (D_a\Psi + Z_a \Psi) (D_b\bar\Psi + Z_b \bar\Psi) + Z_a Z_b \abs{\Psi}^2  \rb]
\ee
which matches the expression in Sec.~5 \cite{PW-em}, and the energy obtained by Gudapati \cite{G-em}.


\bibliographystyle{JHEP}
\bibliography{hertz}       
\end{document}